\documentclass[fleqn,10pt]{wlscirep}
\usepackage[utf8]{inputenc}
\usepackage[T1]{fontenc}
\usepackage{float}
\usepackage{caption}
\usepackage{subcaption}
\usepackage{amsmath,amssymb,amsfonts}
\usepackage{enumitem}
\usepackage{multirow}
\usepackage{algorithm}
\usepackage{algpseudocode}
\newcommand{\cR}{{\mathcal{R}}}
\newcommand{\cC}{{\mathcal{C}}}
\newcommand{\cX}{{\mathcal{X}}}

\newcommand{\cY}{{\mathcal{Y}}}
\newcommand{\cN}{{\mathcal{N}}}
\newcommand{\cM}{{\mathcal{M}}}
\newcommand{\cL}{{\mathcal{L}}}

\newcommand{\cV}{{\mathcal{V}}}

\newcommand{\cE}{{\mathcal{E}}}
\newcommand{\cS}{{\mathcal{S}}}
\newcommand{\bbR}{{\mathbb{R}}}

\newcommand{\bx}{{\mathbf{x}}}
\newcommand{\bz}{{\mathbf{z}}}

\newcommand{\bh}{{\mathbf{h}}}

\newcommand{\bX}{{\mathbf{X}}}
\newcommand{\bY}{{\mathbf{Y}}}

\newcommand{\by}{{\mathbf{y}}}

\newcommand{\bW}{{\mathbf{W}}}

\newcommand{\bC}{{\mathbf{C}}}

\newcommand{\bD}{{\mathbf{D}}}

\newcommand{\bH}{{\mathbf{H}}}

\newcommand{\inner}[1]{\langle #1 \rangle}

\newcommand{\real}{\mathbb{R}}

\newcommand{\map}[3]{#1: #2 \rightarrow #3}

\newtheorem{definition}{Definition}

\newtheorem{lemma}{Lemma}

\title{Renewable-Based Charging in Green Ride-Sharing}

\author[1,+,*]{Elisabetta Perotti}
\author[1,+]{Ana M. Ospina}
\author[2]{Gianluca Bianchin}
\author[3]{Andrea Simonetto}
\author[1]{Emiliano Dall'Anese}
\affil[1]{University of Colorado Boulder, Department of Electrical, Computer, and Energy Engineering, Boulder, USA}
\affil[2]{Université catholique de Louvain, ICTEAM \& Department of Mathematical Engineering, Louvain-la-Neuve, 
Belgium}
\affil[3]{Unité de Mathématiques Appliquées, ENSTA Paris, Institut Polytechnique de Paris, 91120 Palaiseau, France}

\affil[*]{Corresponding author: elisabetta.perotti@colorado.edu}

\affil[+]{These authors contributed equally to this work}

\begin{abstract}
Governments, regulatory bodies, and manufacturers are proposing plans to accelerate the adoption of electric vehicles (EVs), with the goal of reducing the impact of greenhouse gases and pollutants from internal combustion engines on human health and climate change. In this context, the paper considers a scenario where ride-sharing enterprises utilize a $100\%$-electrified fleet of vehicles, and seeks responses to the following key question: How can renewable-based EV charging be maximized without disrupting the quality of the ride-sharing services? We propose a new mechanism to promote EV charging during hours of high renewable generation, and we introduce the concept of \emph{charge request}, which is issued by a power utility company. Our mechanism is inspired by a game-theoretic approach where the power utility company proposes incentives and the ride-sharing platform assigns vehicles to both ride and charge requests; the bargaining mechanism leads to prices and EV assignments that are aligned with the notion of Nash equilibria. Numerical results show that it is possible to shift the EV charging during periods of high renewable generation and adapt to intermittent generation while minimizing the impact on the quality of service. The paper also investigates how the users' willingness to ride-share affects the charging strategy and the quality of service.

\end{abstract}
\begin{document}

\flushbottom
\maketitle

\thispagestyle{empty}

\section*{Introduction}

An increasing portion of the world population is expected to live in urban and sub-urban areas, posing formidable challenges\cite{UN_2019}. The current urban mobility system is obsolete and it requires drastic changes in order to cope with its main flaws such as congestion, inefficiency, and high carbon footprint. Fossil-fuel vehicles are a major contributor to greenhouse gases and pollutants, which in turn are interlinked with climate change and to more than 8 million deaths each year globally\cite{mortalityFossil_2021}. Vehicle electrification is key towards a radically more sustainable mobility system\cite{worldBank_EV_2023}, promising a significant reduction in the environmental impact of the transportation sector. Several ride-sharing and ride-hailing alternatives are emerging in the urban mobility sector, driven by these pressing climate and health-related issues, by increasing traffic congestion, and given the trend of younger generations favoring ride-hailing options over car ownership. Indeed, cities worldwide are already experiencing this transformation, observing the rise in popularity of on-demand ride-hailing options in companies such as Uber and Lyft\cite{ward2021air, GURUMURTHY_2020}. Ride-hailing platforms have started advertising greener options, such as shared rides and electric vehicle (EV) rides, increasing their appeal. Future mobility trends will also include fleets of autonomous EVs for ride-sharing services to improve both the quality of service (QoS) and sustainability\cite{Spieser_2014,BOSCH_2018,HONG_2022}. 

However, a 100\%-electrification of the urban mobility sector -- and, in particular, of the ride-sharing services -- may come at a cost: with the current modus operandi of the power infrastructure, large numbers of EVs may increase the loading of distribution systems, potentially surpassing the loading capacity in portions of the grid\cite{paudyal2021ev,panossian2022challenges}; this, in turn, would compromise the reliability and increase the fragility of the power infrastructure\cite{worldBank_EV_2023}. Moreover, large swings in the power demand from EVs may impact electricity prices and energy markets. It is therefore of paramount importance to uncover coordination mechanisms between the power network operators and transportation system operators to enable a reliable and effective integration of EVs into the grid at large scale\cite{Lingwen_2013,Rossi_2020}. 

In this context, we consider a scenario where ride-hailing enterprises utilize a $100\%$-electrified fleet of vehicles, and seek responses to the following key questions: given the potential effect of EV charging on the power infrastructure, how can power utility companies and ride-hailing companies interact to promote vehicle charging in areas with high renewable generation? How can one maximize the utilization of renewables for charging purposes without disrupting the quality of the ride-hailing services? Answering the first question would allow one to systematically integrate EVs at scale with minimal effect on the power grid reliability, and without requiring structural upgrades of distribution feeders and substations to handle the additional power demand. A positive response to the second question would provide evidence for a successful transition away from internal combustion engines in the urban mobility sector. In this paper, we provide answers to the questions above by investigating new means for power utility companies to interact with ride-hailing companies in order to promote renewable-based charging directly at locations where renewables are available. Beyond maximizing the use of renewable generation, the aim is also to satisfy the largest number of ride requests and to keep the unoccupied fleet size as small as possible. 

Before describing the proposed methodology, we provide a brief overview of existing approaches in the context of mobility-as-a-service (MaaS) and for the coordination between transportation and power systems. Various approaches to tackle problems related to the dispatch of ride-hailing fleets can be found in the literature, including microscopic (possibly stochastic) combinatorial problems\cite{AS-LM-CG:19, Dafernos_1969, JA-SS-AW-EF-DR:17, beirigo2022business,bongiovanni2022machine,fielbaum2022anticipatory} and macroscopic network-flow-based formulations\cite{BT-MA:20}. A common topic of research is the interaction between a large fleet of (autonomous) EVs and the power infrastructure in densely populated areas, taking into account factors such as EV charging requirements\cite{Zhang_2020}, fluctuating customer demand, battery degradation, and power system constraints\cite{Rossi_2020, BT-MA:20, brandstatter2016overview, Mahnoosh_2017, Spieser_2016}. It is also well acknowledged that EVs have the potential to benefit the grid by providing\cite{Mouli_2019, EV_to_grid2, Pengcheng_2022}: (i) energy storage, serving as distributed power storage unit storing excess energy generated by renewable resources; (ii) load balancing, scheduling the EV charging during off-peak hours; and (iii) ancillary services, such frequency support. 

The coupled problem arising from a transportation network together with a power network load balancing has been analyzed via single-level optimization formulations\cite{Mahnoosh_2017}, exploiting the benefits of EV charging with renewable energy. Some works have explored hierarchical models such as Nash–Stackelberg–Nash game framework in a network-flow based formulation\cite{Lv_2023, Wenjie_2022, EV_to_grid1, EV_RES7, EV_RES5, EV_RES1}. Traditional game-theoretic frameworks have analyzed the interaction between EVs and the grid through the regulation of energy prices and EV charging schedules, where the grid acts as a player that sets the cost of the energy, and EVs respond with a charging schedule that optimizes their operation, usually assuming traffic flow models\cite{Yoshihara_2018}. Other lines of work model the problem of coordinating the charging needs of an EV fleet as a game, seeking a Nash equilibrium\cite{Ma_2013}. In the case where the EV charging management problem is affected by the volatile nature of renewable generation, Generalized Nash Equilibrium (GNE) approaches have been used to coordinate the EV charging plans with real-time generation profiles\cite{Wei_2016, Chen_2020}. 

\begin{figure}[t!] 
\captionsetup{font={small,it},labelfont={bf,sf}}
\captionsetup[sub]{font=footnotesize,labelfont={}}
\centering
\begin{subfigure}{.45\textwidth} \centering
  \includegraphics[width=\textwidth]{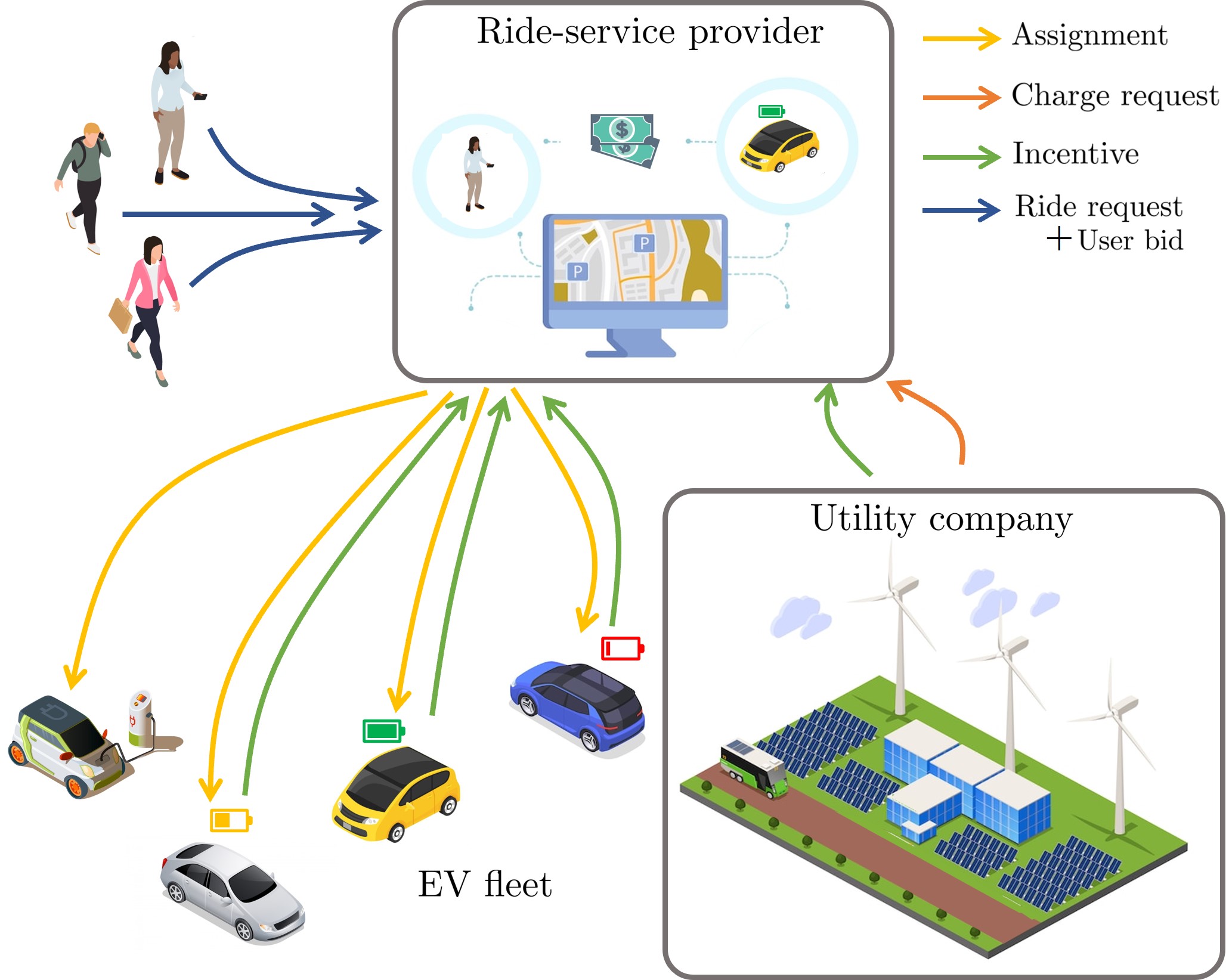}
  \caption{}\label{fig:cartoonA}
\end{subfigure}
\centering
\begin{subfigure}{.40\textwidth} \centering
  \includegraphics[width=1.6in]{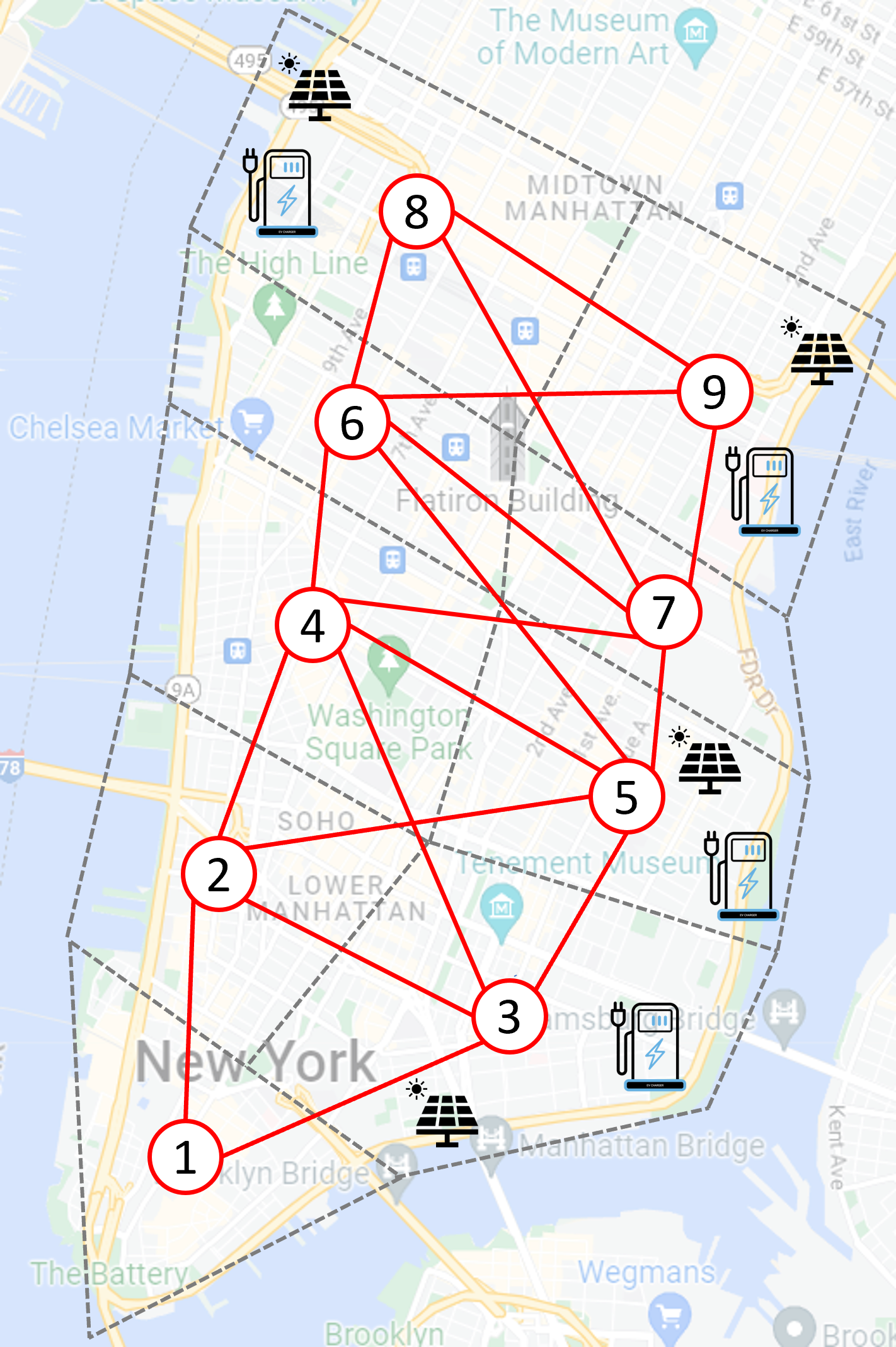}
  \caption{} \label{fig:cartoonB}
\end{subfigure}
\caption{(a) Description of the interaction between utility company, ride-service provider, and EV drivers. The ride-service provider receives ride requests from customers and charge requests from a utility company. The bargaining procedure involves a vehicle-request assignment problem that takes into consideration incentives coming from both parties and bids from customers, to assign available EVs to requests. (b) Case study: Lower Manhattan, New York City, NY, partitioned into 9 regions (links between regions correspond to a reachable site within a 10 minute drive). PV generation is present in 4 regions. This image has been designed using assets from Freepik.com.}
\label{fig:cartoons}
\vspace{-.4cm}
\end{figure}

In this paper, we propose a new mechanism to enable interactions between power utility companies and ride-hailing companies. We also consider the case where rides may be shared, namely, multiple ride requests can be served by the same vehicle. The proposed process requires minimal modifications of existing vehicle-ride assignment frameworks for ride-hailing services -- where rides are assigned to vehicles based on an assignment problem -- and with a little computational and operational burden on the power utility side. 
The proposed mechanism enables a power utility company to issue \emph{charge requests} that model a financial incentive (tied to specific renewable generation profiles and locations) offered to the ride-hailing company to promote the use of the available renewable energy. This mechanism is qualitatively illustrated in Figure \ref{fig:cartoonA}. Drawing from game-theoretic approaches, our strategy involves a bargaining procedure where the power company proposes incentives, and the ride-hailing platform, after receiving charging incentives and ride requests from customers, assigns vehicles to both ride requests and charge requests.
Mathematically, the interaction is in the form of a Gauss-Seidel method where, at each iteration, the power utility company proposes new incentives associated with the charge requests by solving a given optimization problem, and based on the current potential vehicle assignment; subsequently, the ride-service provider issues a new potential assignment based on the new incentives. The performance cost used in the vehicle-request assignment problem involves the minimization of the overall operational cost for the ride-service provider. The cost used by the utility company quantifies the need for the self-consumption of renewable energy resources. With this mechanism, the power utility company and the ride-service provider eventually lead to prices and EV assignments that are aligned with the notion of Nash equilibria. We point out that, while this work stresses the renewable generation profile, the same bargaining procedure can be utilized by the utility for general desirable power demand profiles; this opens the door to setups where the ride-hailing company acts as a virtual power plant (VPP) providing services to the grid at convenient financial conditions. 

Our approach retains the microscopic feature of existing assignment problems in the context of MaaS platforms~\cite{li2017mobility,butler2021barriers,hensher2020understanding}; moreover, following the traditional workflow with limited information sharing, it can be integrated seamlessly into MaaS orchestration services (indeed, as shown in Figure \ref{fig:cartoonA}, the utility company can be understood just as an additional ``entity'' into a vehicle assignment task). At the same time, it naturally models and enables an interaction between power systems operations and EV-dominated ride-service providers without resorting to macroscopic (or averaged) flow models, which would be difficult to integrate into existing fleet assignment frameworks. Another important feature of our approach is that it does not rely upon combinatorial problems (therefore, it does not require dedicated software), and can leverage simple optimization algorithms. We test the proposed mechanism and show that it is indeed possible to shift the EV charging during periods of high renewable generation and adapt to intermittent generation. Our approach does not cause a degradation of the QoS for the ride-service provider with respect to ad hoc EV charging strategy.

\section*{Results}

The following results stem from the mathematical algorithms described in the Methods section. The experiments consider a fleet of 100 EVs and use the data recorded by the Taxi and Limousine Commission (TLC)\cite{TLC} on Tuesday, March 1, 2022, between 6:00 and 24:00. In terms of renewable generation, we consider the photovoltaic (PV) power generation profiles extracted from the renewable historical data of New York Independent System Operator\cite{NY_ISO_2023}, as explained in the Data Sources section. To clarify, we use the term ride-hailing below to indicate the case where each vehicle carries only one passenger, as in a taxi service. On the other hand, we use the term ride-sharing to refer to the case where several riders can be picked up along a route by the same vehicle. In line with this terminology, even shared/pooled ride-hailing services are included in this category. Note that both ride-sharing and ride-hailing platforms are more generically referred to as ride-service providers. The complete description of the simulation setup is provided in the Methods section. To present the results we define three study cases:

\noindent $\bullet$ \emph{Business-as-usual}: the ride-hailing platform assigns ride requests to an EV fleet. The EVs are exclusively charging when their batteries are empty. The EV charging does not exploit renewable energy resources.

\noindent $\bullet$ \emph{Case 1}: the ride-hailing platform receives and assigns both ride and charge requests, based on the outcome of the process described in Methods, in which the EV fleet and the power utility company engage in an incentive-assignment process. 

\noindent $\bullet$ \emph{Case 2}: this extends \emph{case 1} enabling the ride-sharing option: here, EVs 
are allowed to pick up more passengers during the trip provided they are willing to ride-share and have a 
common destination. Each additional passenger can cause a delay of up to 4 minutes.

\begin{figure}[H]
\captionsetup{font={small,it},labelfont={bf,sf}}
\captionsetup[sub]{font=footnotesize,labelfont={}}
\centering
  \includegraphics[width=0.85\textwidth]{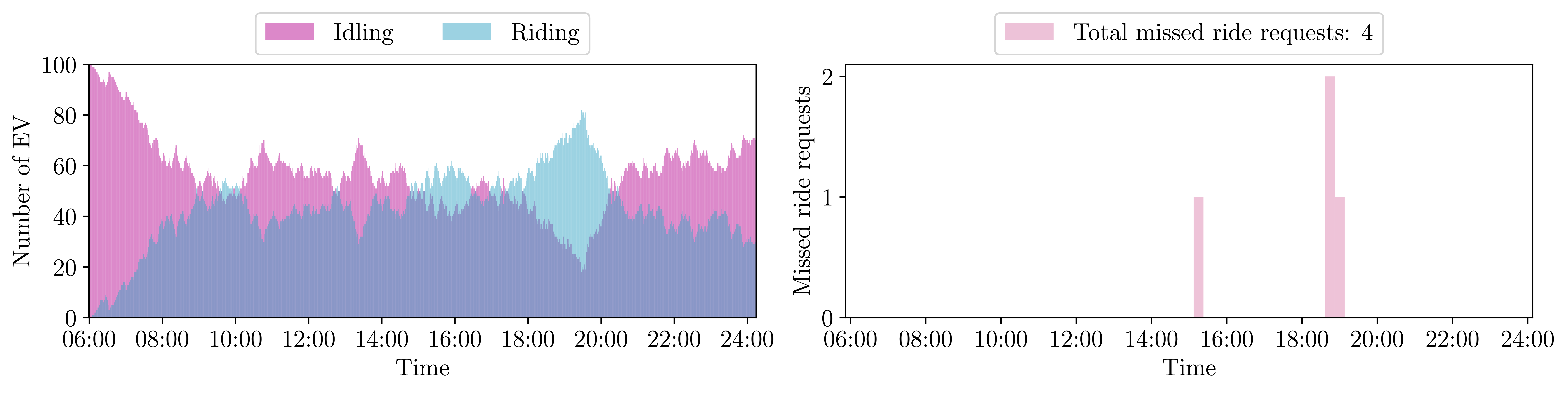}
\caption{Availability of fossil-fuel vehicles during the day (left) and total number of missed ride requests (right). With a fleet of only fossil-fuel vehicles, the QoS is 99.8\%.}
\label{fig:fossil_fuel}
\end{figure}

\emph{Case 1} and \emph{case 2} are evaluated under three different weather conditions, named sunny day, cloudy morning, and cloudy afternoon, that correspond to different PV generation profiles. To assess the performance of the proposed strategy in the cases defined above, we introduce two metrics. These are: the \emph{Quality of Service (QoS)}, defined as the percentage of attended ride requests out of the total number of received ride requests, and the \emph{Power Loss (PL)}, defined as the percentage of unexploited renewable power out of the total PV power generated. A summary of the results is presented in Table \ref{tab:metrics}. Notice how the QoS always improves with the sharing acceptance, e.g., going from 94.8\% to 99.9\% on a sunny day, indicating that ride-sharing is critical to enforce high levels of QoS. On the other hand, the amount of unexploited renewable power increases as the willingness to ride-share improves. In this case, the EVs need to charge less often since their total driving time is overall reduced, given the same (or larger) number of accepted ride requests.

To provide some context, we start by considering a fleet consisting of regular fossil-fuel vehicles. Figure \ref{fig:fossil_fuel} shows the availability of vehicles over the simulation window, where if an EV is attending a ride request it is labeled as ``riding'', otherwise, if it is unoccupied it is labeled as ``idling''. In the Discussion, we will compare the performance of a fossil-fuel fleet against a 100\% EV fleet, given the same amount of received ride requests.

\begin{table}[H]
\captionsetup{font={small,it},labelfont={bf,sf}}
\captionsetup[sub]{font=footnotesize,labelfont={}}
\centering
\begin{tabular}{|c|c||c c||c c| c c| c c|c c||}
\cline{3-12}
\multicolumn{1}{c}{} & & \multicolumn{2}{|c||}{\textbf{Case 1}} & \multicolumn{8}{|c||}{\textbf{Case 2}} \\
\cline{5-12}
\multicolumn{1}{c}{} & & \multicolumn{2}{|c||}{} & \multicolumn{2}{|c|}{$\mathbf{100\%}$} & \multicolumn{2}{|c|}{$\mathbf{75\%}$} & \multicolumn{2}{|c|}{$\mathbf{50\%}$} & \multicolumn{2}{|c||}{$\mathbf{25\%}$} \\
\cline{2-12}
\multicolumn{1}{c|}{} & \textbf{Metric} & QoS & PL & QoS & PL & QoS & PL & QoS & PL & QoS & PL \\
\hline
\hline
\multirow{3}{*}{\textbf{Weather}} & Sunny day & 94.8\% & 36.7\% & 99.9\% & 45.0\%  & 98.9\% & 41.7\% & 97.6\% & 39.6\% & 95.7\% & 37.3\% \\
& Cloudy morning &  93.6\% & 19.8\% & 100\% & 24.6\%  &  99.6\% & 21.6\% & 97.3\% & 20.7\% & 94.6\% & 20.1\%  \\
& Cloudy afternoon & 93.9\% & 24.3\% & 99.5\% & 28.1\% &  98.1\% & 26.2\% & 96.5\% & 25.2\% & 94.6\% & 24.5\% \\
\hline
\end{tabular}
\caption{\label{tab:metrics} Quality of service (QoS), and power loss (PL) evaluated over the simulation window. In \emph{case 1}, the ride-service provider assigns ride and charge requests, and the EV fleet and utility company implement the bargaining procedure described in the Methods section. \emph{Case 2} adds the ride-sharing option to \emph{case 1}; here we assume that 100\%, 75\%, 50\%, and 25\% of the customers are willing to ride-share. As a benchmark metric, the QoS for the \emph{business-as-usual} case is 94.5\%.}
\end{table}

\subsection*{Effects of business-as-usual EV charging on the power infrastructure} 
\begin{figure}[t!]
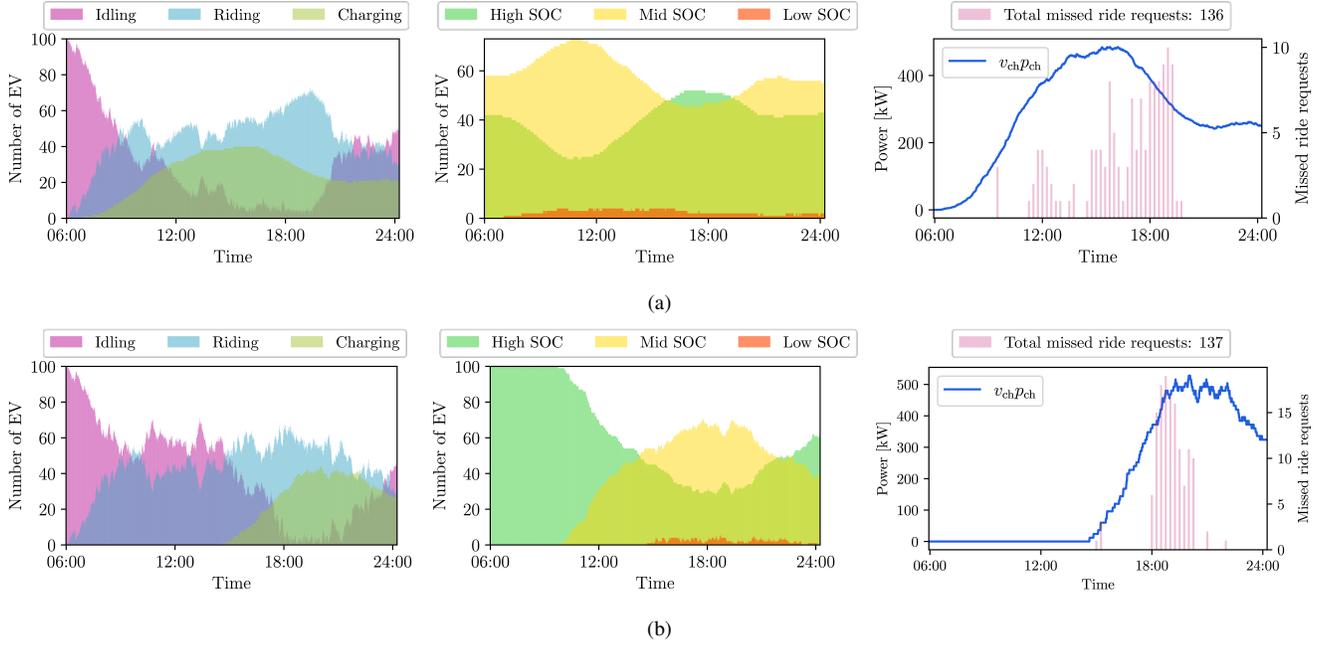

\captionsetup{font={small,it},labelfont={bf,sf}}
\captionsetup[sub]{font=footnotesize,labelfont={}}
\centering
\begin{subfigure}{1\textwidth} \centering
  \includegraphics[width=\textwidth]{NoChargeReq_avg_rev.png}
  \caption{}
  \label{fig:NoChargeReq_SOC}
\end{subfigure}
\centering
\begin{subfigure}{1\textwidth} \centering
  \includegraphics[width=\textwidth]{NoChargeReq_100_rev.png}
  \caption{}
  \label{fig:NoChargeReq_100}
\end{subfigure}
\caption{Availability of EVs during the day, SOC time-evolution, and charging profiles together with total number of missed ride requests given two different initial SOC conditions: random distribution between $10\%$ and $100\%$ of the battery capacity (a) and fully charged (b). Business-as-usual case. Dark blue line corresponds to the charging profile (i.e., total power used to charge $v_{\mathrm{ch}}$  EVs, each receiving $p_{\mathrm{ch}}$ kW).}
\label{fig:NoChargeReq}
\end{figure}

Initially, we consider a vehicle-request assignment where the ride-service provider has to handle ride requests only. This case, which disregards the presence of renewable energy resources, will serve as a benchmark for later comparison, and represent the \emph{business-as-usual} case.  
The assignment is performed as described in, e.g.,~\cite{AS-LM-CG:19}, for the available EVs. We assume that the EVs will start charging when their State Of Charge (SOC) is low ($<10\%$ of the battery capacity), regardless of their location and the time of the day. Figure \ref{fig:NoChargeReq} shows the availability of EVs, the SOC trend during the day, and the EV charging profile (i.e., the amount of power used to charge $v_{\mathrm{ch}}$  EVs, at a charging rate $p_{\mathrm{ch}}$), with the number of missed ride requests. For convenience, we consider three levels of SOC: high ($>60\%$), mid (between $10\%$ and $60\%$), and low ($<10\%$). We explore two different initial SOC conditions: random and fully charged. For random SOC, we present averaged results over 100 iterations. In this case, without charge requests, the EVs connect to the power grid whenever they are out of battery, generating a peak of power consumption. We show the availability of EVs over the day for both cases in Figure \ref{fig:NoChargeReq}, where if an EV is busy charging it is labeled as ``charging''. In Figure \ref{fig:NoChargeReq_SOC}, where the initial SOC of EVs are randomly generated, the peak of the EV charging profile occurs between 13:00 and 17:00, while in Figure \ref{fig:NoChargeReq_100}, where all the EVs start fully charged at 6:00, the peak is shifted, starting at 20:00. In both cases, we observe that the majority of missed ride requests occurs in the time interval between 18:00 and 20:00.

\subsection*{Renewable-based charging and bargaining strategy} 
\begin{figure}[t!]
\captionsetup{font={small,it},labelfont={bf,sf}}
\captionsetup[sub]{font=footnotesize,labelfont={}}
\centering
\begin{subfigure}{.32\textwidth}
  \centering
  \includegraphics[width=\textwidth]{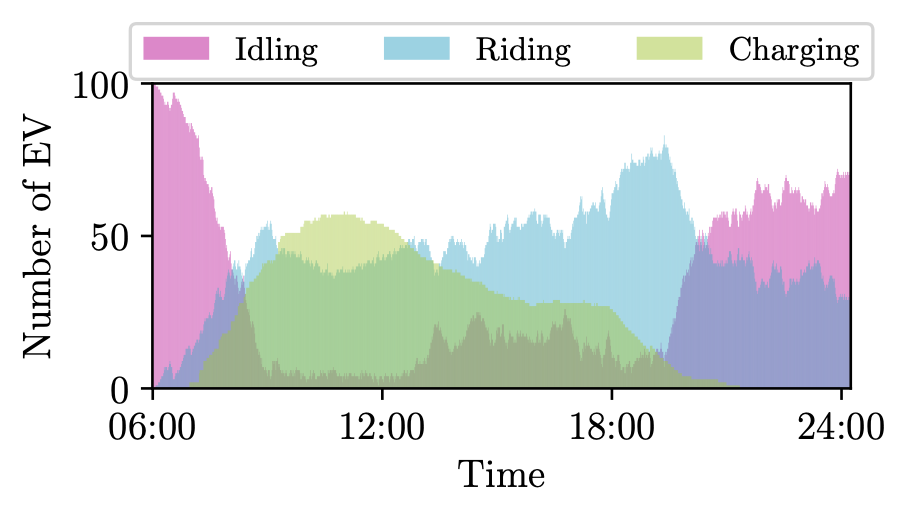}
  \caption{}
  \label{fig:BiSu_EV_avg}
\end{subfigure}
\centering
\begin{subfigure}{.32\textwidth}
  \centering
  \includegraphics[width=\textwidth]{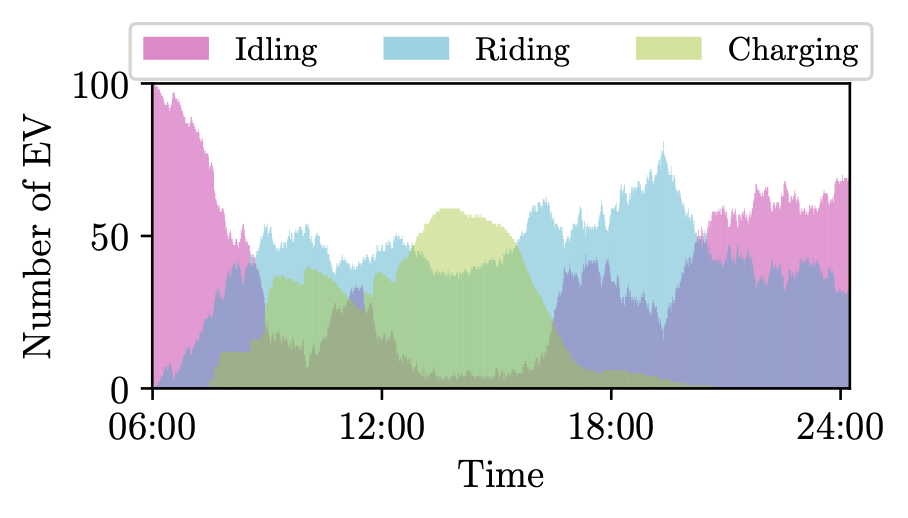}
  \caption{}
  \label{fig:BiAM_EV_avg}
\end{subfigure}
\centering
\begin{subfigure}{.32\textwidth}
  \centering
  \includegraphics[width=\textwidth]{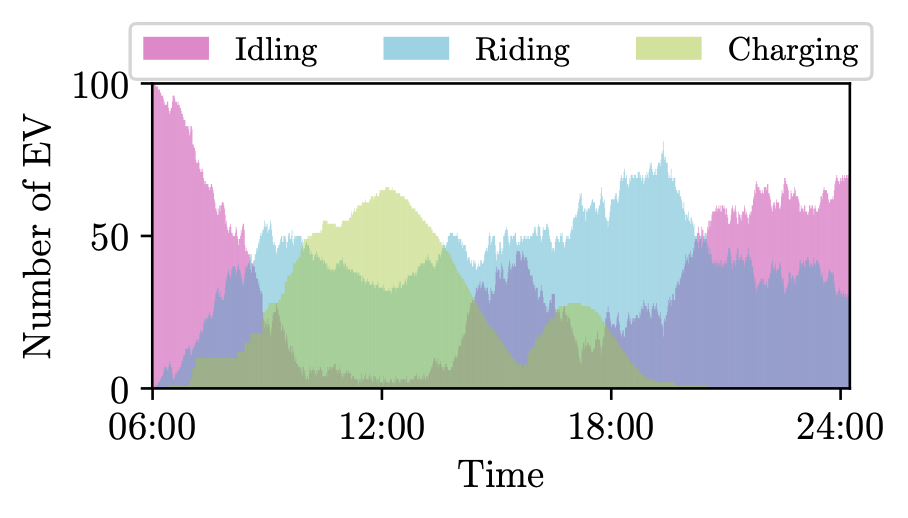}
  \caption{}
  \label{fig:BiPM_EV_avg}
\end{subfigure}
\centering
\begin{subfigure}{.32\textwidth}
  \centering
  \includegraphics[width=\textwidth]{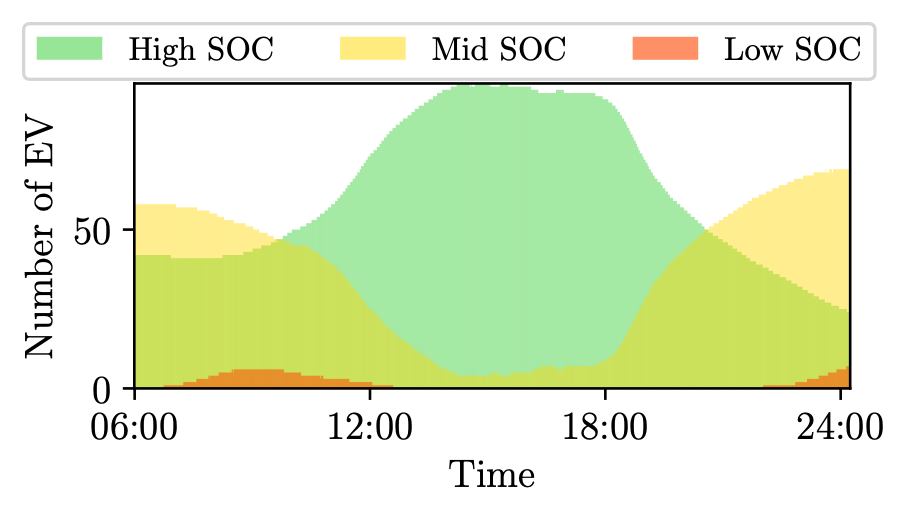}
  \caption{}
  \label{fig:BiSu_SOC_avg}
\end{subfigure}
\centering
\begin{subfigure}{.32\textwidth}
  \centering
  \includegraphics[width=\textwidth]{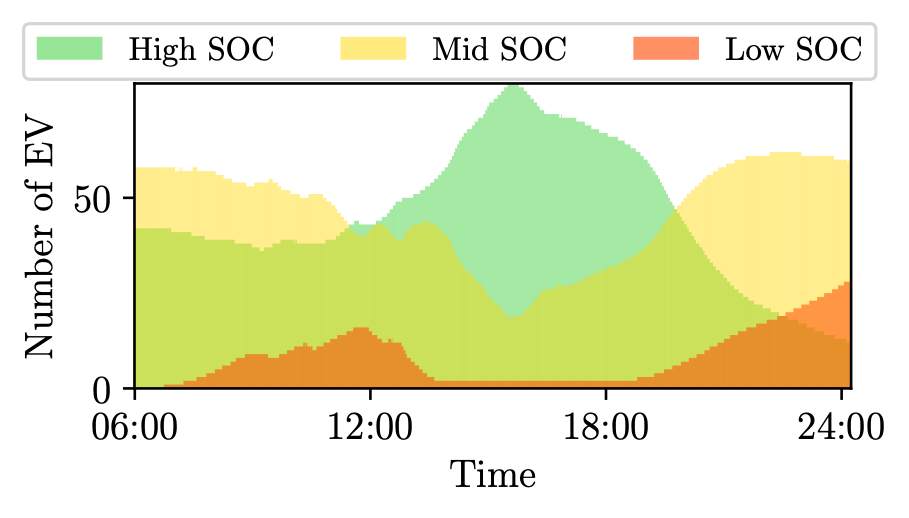}
  \caption{}
  \label{fig:BiAM_SOC_avg}
\end{subfigure}
\centering
\begin{subfigure}{.32\textwidth}
  \centering
  \includegraphics[width=\textwidth]{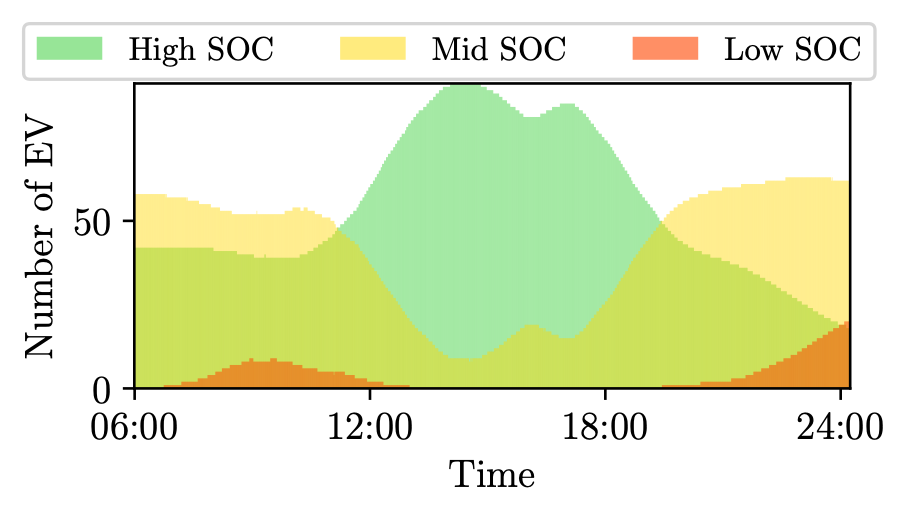}
  \caption{}
  \label{fig:BiPM_SOC_avg}
\end{subfigure}
\centering
\begin{subfigure}{.32\textwidth}
  \centering
  \includegraphics[width=\textwidth]{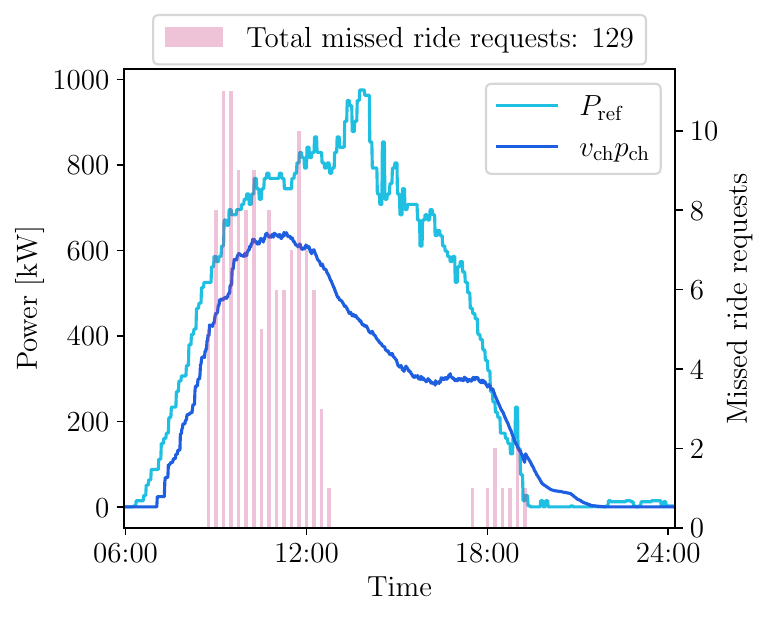}
  \caption{}
  \label{fig:BiSu_PrefMissRide_avg}
\end{subfigure}
\centering
\begin{subfigure}{.32\textwidth}
  \centering
  \includegraphics[width=\textwidth]{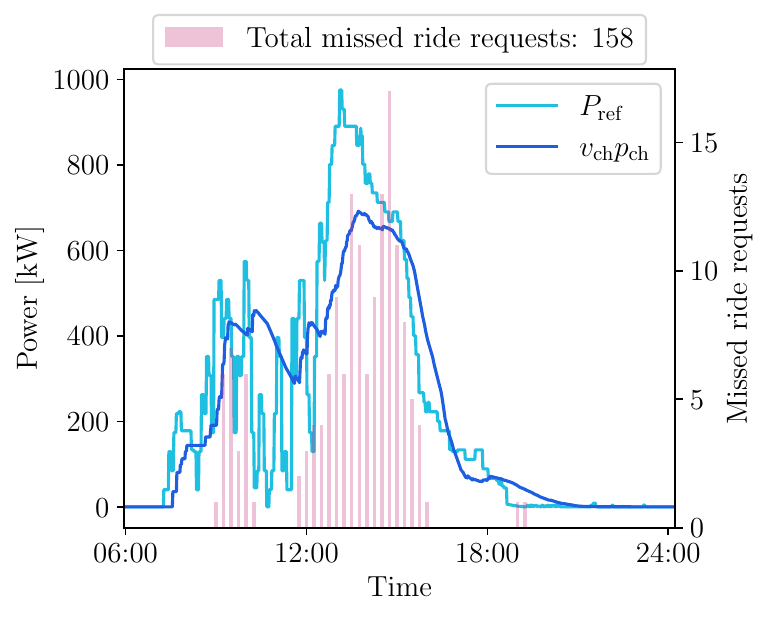}
  \caption{}
  \label{fig:BiAM_PrefMissRide_avg}
\end{subfigure}
\centering
\begin{subfigure}{.32\textwidth}
  \centering
  \includegraphics[width=\textwidth]{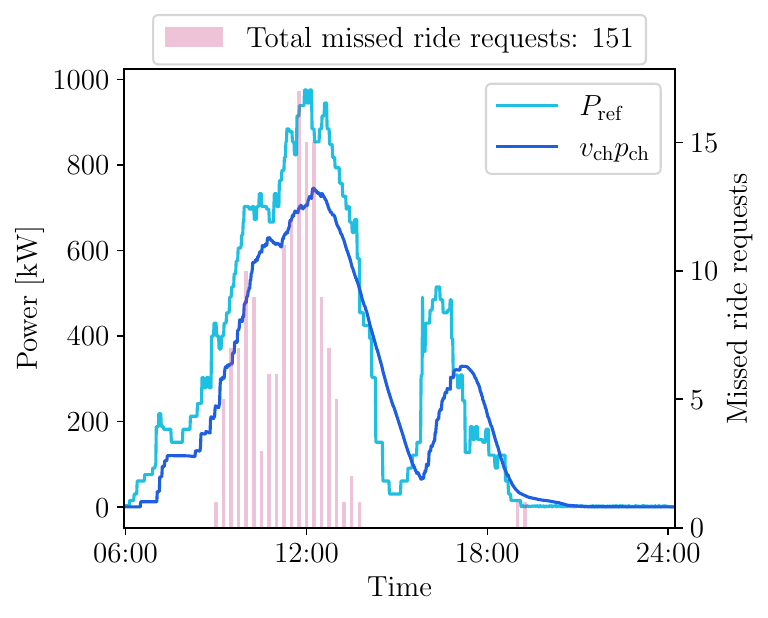}
  \caption{}
  \label{fig:BiPM_PrefMissRide_avg}
\end{subfigure}
\centering
\begin{subfigure}{0.32\textwidth}
  \centering
  \includegraphics[width=\textwidth]{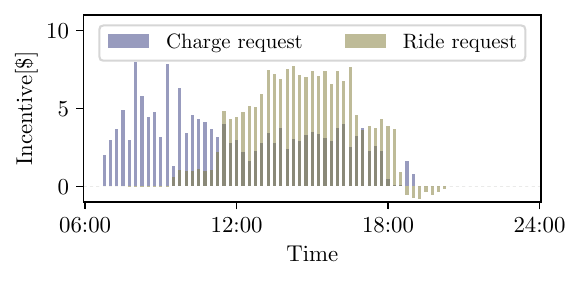}
  \caption{}
  \label{fig:BiSu_incAssign_avg}
\end{subfigure}
\centering
\begin{subfigure}{0.32\textwidth}
  \centering
  \includegraphics[width=\textwidth]{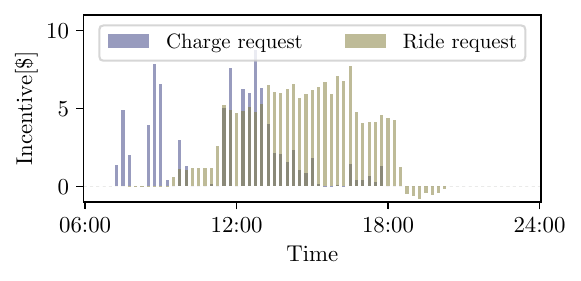}
  \caption{}
  \label{fig:BiAM_incAssign_avg}
\end{subfigure}
\centering
\begin{subfigure}{0.32\textwidth}
  \centering
  \includegraphics[width=\textwidth]{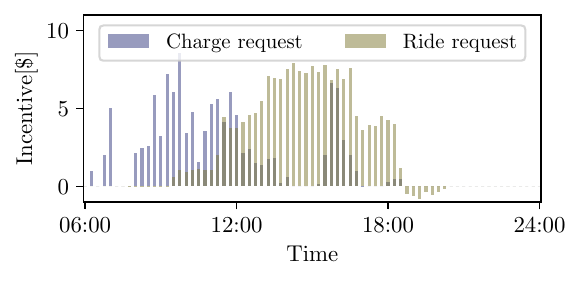}
  \caption{}
  \label{fig:BiPM_incAssign_avg}
\end{subfigure}
\caption{Availability of EVs during the day ($1^{\mathrm{st}}$-row), SOC time-evolution ($2^{\mathrm{nd}}$-row), charging profiles together with the total number of missed ride requests ($3^{\mathrm{rd}}$-row) and incentives ($4^{\mathrm{th}}$-row) during the day, given three different weather scenarios: sunny (a)-(d)-(g)-(j), cloudy morning (b)-(e)-(h)-(k), and cloudy afternoon (c)-(f)-(i)-(l). The initial SOC of each EV is randomly set within $10\%$ and $100\%$ of the battery capacity. The PV-generation ($P_{\mathrm{ref}}$) and charging profiles ($v_{\mathrm{ch}} p_{\mathrm{ch}}$) are obtained summing over all charging stations.}
\label{fig:Bilevel}
\end{figure}

\begin{figure}[t!]
\captionsetup{font={small,it},labelfont={bf,sf}}
\captionsetup[sub]{font=footnotesize,labelfont={}}
\centering
\begin{subfigure}{1\textwidth}
  \includegraphics[width=\textwidth]{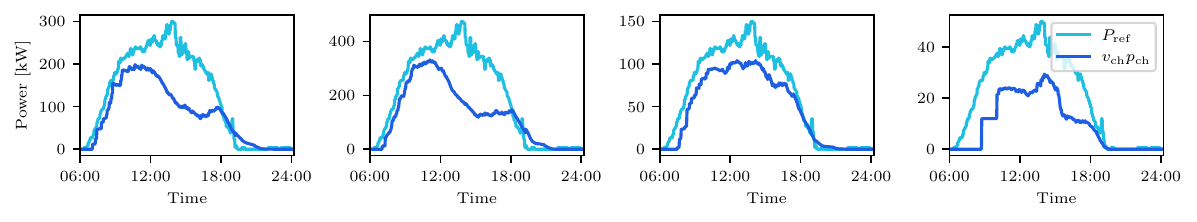}
  \caption{}
  \label{fig:BiSu_power4stations_avg}
\end{subfigure}
\centering
\begin{subfigure}{1\textwidth}
  \includegraphics[width=\textwidth]{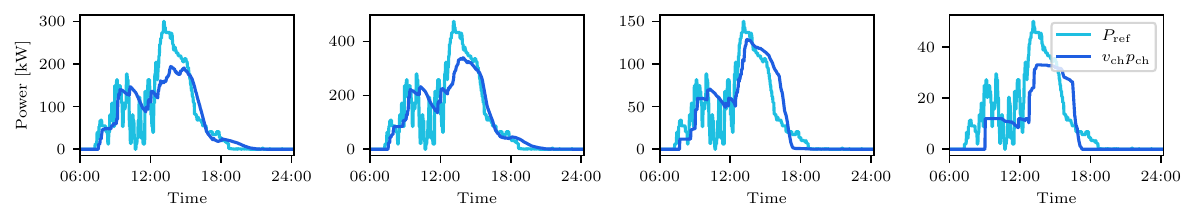}
  \caption{}
  \label{fig:BiAM_power4stations_avg}
\end{subfigure}
\centering
\begin{subfigure}{1\textwidth}
  \includegraphics[width=\textwidth]{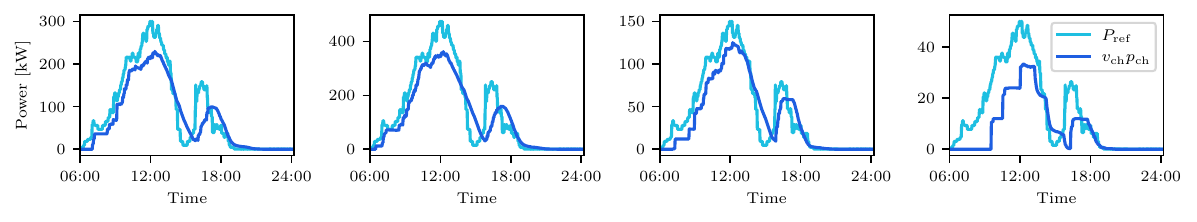}
  \caption{}
  \label{fig:BiPM_power4stations_avg}
\end{subfigure}
\caption{Charging profile for each of the 4 regions hosting charging facilities, corresponding (left to right) to node 3, 5, 8 and 9 in Figure \ref{fig:cartoonB}, given three different weather scenarios: sunny (a), cloudy morning (b), and cloudy afternoon (c). The curves in dark and light blue correspond to the PV-generation ($P_{\mathrm{ref}}$) and the charging profiles ($v_{\mathrm{ch}} p_{\mathrm{ch}}$), respectively.}
\label{fig:Bilevel_power4sation}
\vspace{-.3cm}
\end{figure}

We focus now on \emph{case 1}, where the ride-service provider handles both ride and charge requests. In this case, the ride-hailing company, the power utility company, and  the EVs interact as described in the Methods section and as illustrated in Figure~\ref{fig:cartoonA}. We recall that charge requests are issued by the power utility company. EVs are assigned to feasible requests, depending on their distance to the customers' pickup point or to the charging facilities. The shortest path is assumed to correspond to the lower cost and therefore preferred by the ride-hailing company. Moreover, a crucial aspect of our approach is to rely on a bargaining mechanism, in the form of financial incentives, that influences the preferences of the ride-hailing platform concerning the assignment of EVs to ride or charge requests. In the bargaining mechanism (a rigorous formulation of the problem can be found in the Methods section), we augment a linear assignment problem with additional constrained optimization problems, whose objective is to assign a financial incentive to the cost of the ride or charge requests based on the bids from the users and the losses affecting the power utility company. We present averaged results over 100 Monte Carlo simulations, where each EV is initially assigned a random SOC between $10\%$ and $100\%$ of the full battery capacity. In Figure \ref{fig:Bilevel}, the top row shows the availability of EVs through the simulation window; the second row displays how the SOC of the EV fleet varies during the day under different weather conditions; the third row shows the total PV generation profile in the lower Manhattan area, compared to the power distributed to charging EVs; and the fourth row presents the incentives for ride and charge requests. Notice how the EV charging schedule now follows the PV generation profile and it is impacted by the weather conditions. Figure \ref{fig:Bilevel_power4sation} shows the individual PV generation profiles and the amount of power dispatched in each of the four areas equipped with charging facilities. Then, our method successfully can shift the charging profiles to periods of the day characterized by high renewable generation and avoid the negative impacts on the grid that the wider charging schedule showed in Figure \ref{fig:NoChargeReq} can induce.

Regarding the effects of the variability of renewable energy resources on our formulation, we can see in Figures \ref{fig:BiAM_SOC_avg}-\ref{fig:BiPM_SOC_avg} the impact on the SOC for two different PV profiles, both affected by significant cloud coverage. In the case of a cloudy morning, EVs cannot fully charge between 9:00 and 13:00, resulting in a larger amount of missed ride requests during the rest of the day (see Figure \ref{fig:BiAM_PrefMissRide_avg}), and many more EVs with low SOC by midnight, as compared to the sunny day case. In contrast, the cloudy afternoon profile still allows many EVs to fully charge in the morning, substantially limiting the need for charging in the afternoon and therefore affecting less the final SOC of the EVs. Still, more ride requests are missed with respect to the sunny day, as shown in Figure \ref{fig:BiPM_PrefMissRide_avg}. Note that Figure \ref{fig:BiAM_incAssign_avg} shows fewer charge request incentives from 8:00 to 12:00, due to a drop in the PV generation in the morning. In the cloudy afternoon scenario, however, during the same period, the charge request incentives are higher, and therefore, the EV charging is favored over attending ride requests as shown in Figure \ref{fig:BiPM_incAssign_avg}. The charge request financial incentives depend on the PV profile scenario considered, i.e., sunny days are characterized by a considerable amount of incentives during the day following the consistent PV-power generation, while cloudy days experience a drop in the amount of charge request incentives during the morning and afternoon, respectively.

\subsection*{The benefits of ride-sharing}
We explore the impact of ride-sharing in \emph{case 2}, by allowing EVs to pick up more passengers provided that they share a common destination. Passengers can be picked up along the journey, i.e., they do not need to start from the same pick-up area. It is important to stress that we are still dealing with a one-to-one vehicle-request assignment, while we keep track of the number of passengers on board which can be at most 4. Passengers express their willingness to ride-share  whenever they submit a  request, in exchange for a discounted price and a greater chance to get a ride.

\begin{figure}[t!]
\captionsetup{font={small,it},labelfont={bf,sf}}
\captionsetup[sub]{font=footnotesize,labelfont={}}
\centering
\begin{subfigure}{.32\textwidth}
  \centering
  \includegraphics[width=\textwidth]{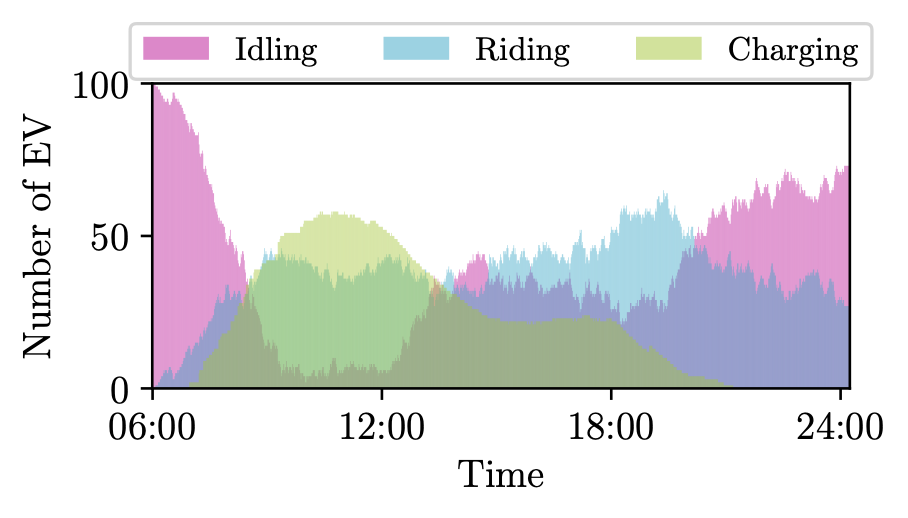}
  \caption{}
  \label{fig:75Su_EV_avg}
\end{subfigure}
\centering
\begin{subfigure}{.32\textwidth}
  \centering
  \includegraphics[width=\textwidth]{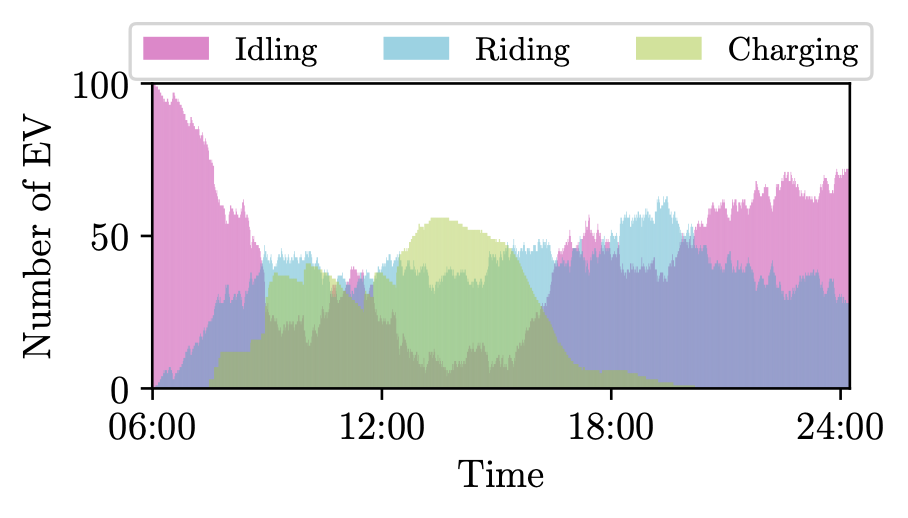}
  \caption{}
  \label{fig:75AM_EV_avg}
\end{subfigure}
\centering
\begin{subfigure}{.32\textwidth}
  \centering
  \includegraphics[width=\textwidth]{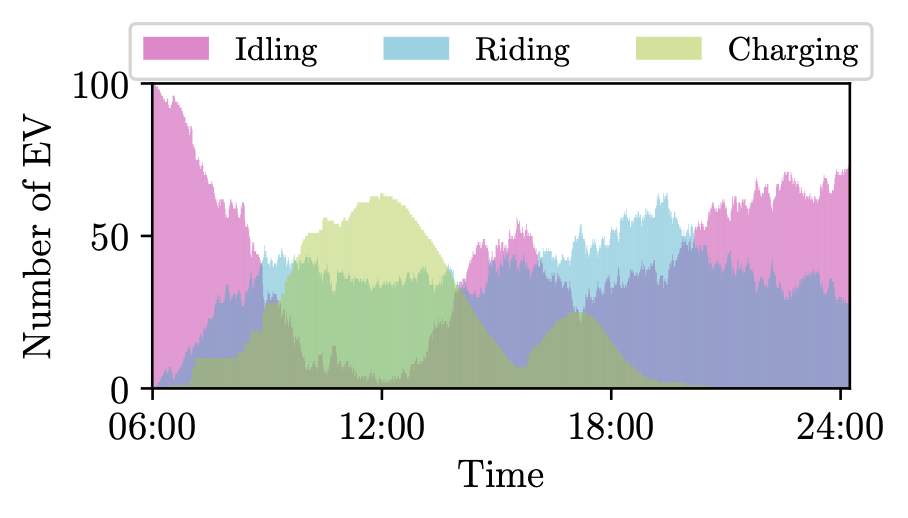}
  \caption{}
  \label{fig:75PM_EV_avg}
\end{subfigure}
\centering
\begin{subfigure}{.32\textwidth}
  \centering
  \includegraphics[width=\textwidth]{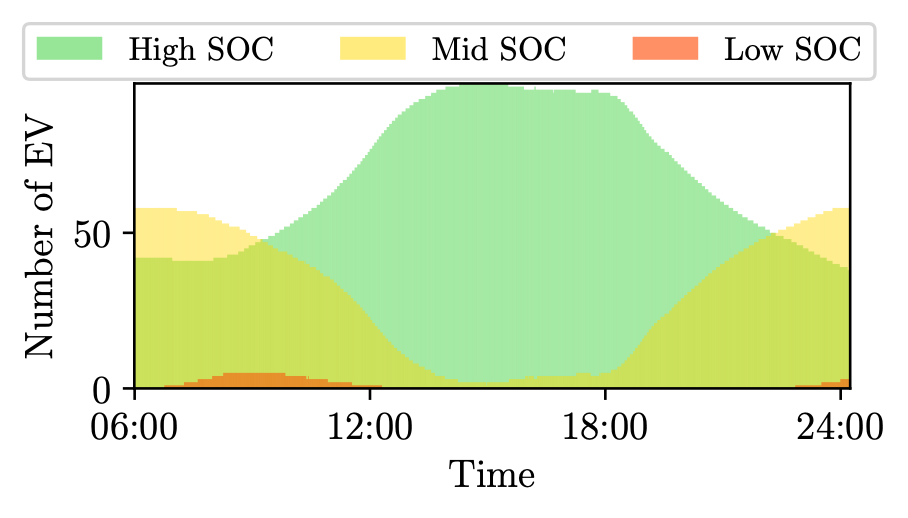}
  \caption{}
  \label{fig:75Su_SOC_avg}
\end{subfigure}
\centering
\begin{subfigure}{.32\textwidth}
  \centering
  \includegraphics[width=\textwidth]{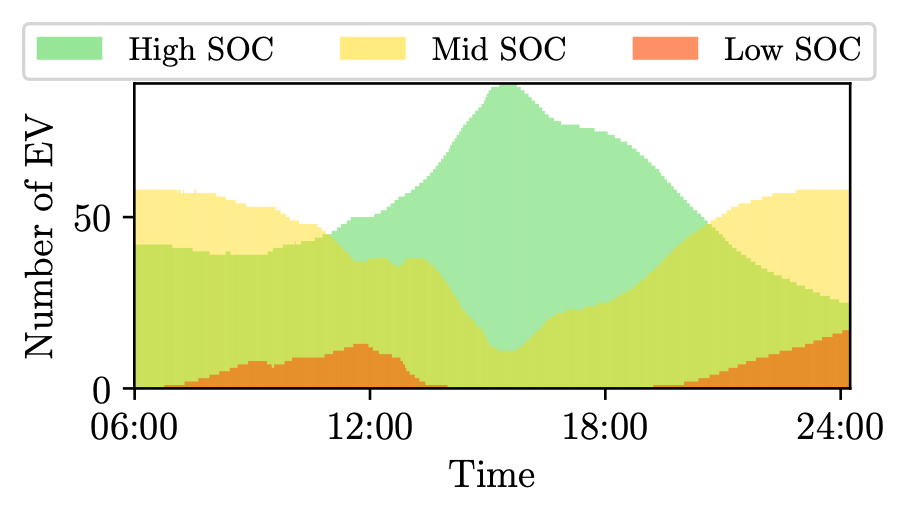}
  \caption{}
  \label{fig:75AM_SOC_avg}
\end{subfigure}
\centering
\begin{subfigure}{.32\textwidth}
  \centering
  \includegraphics[width=\textwidth]{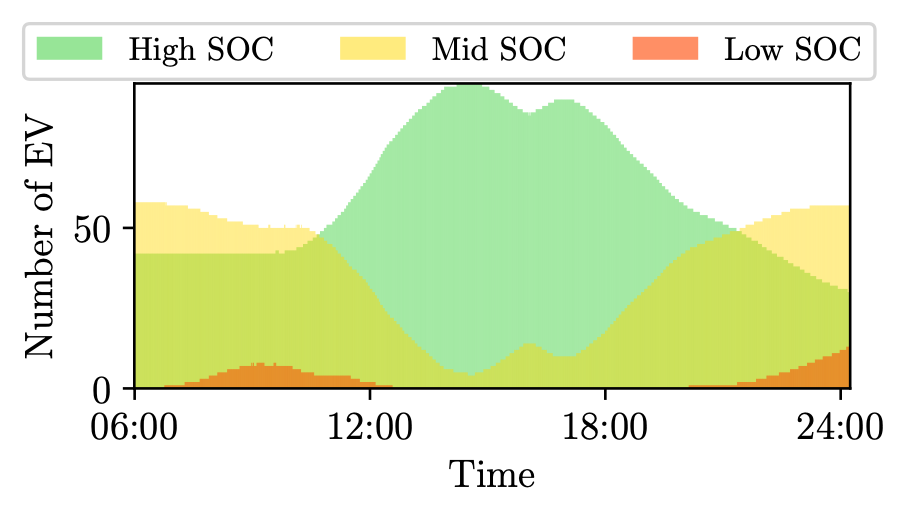}
  \caption{}
  \label{fig:75PM_SOC_avg}
\end{subfigure}
\centering
\begin{subfigure}{.32\textwidth}
  \centering
  \includegraphics[width=\textwidth]{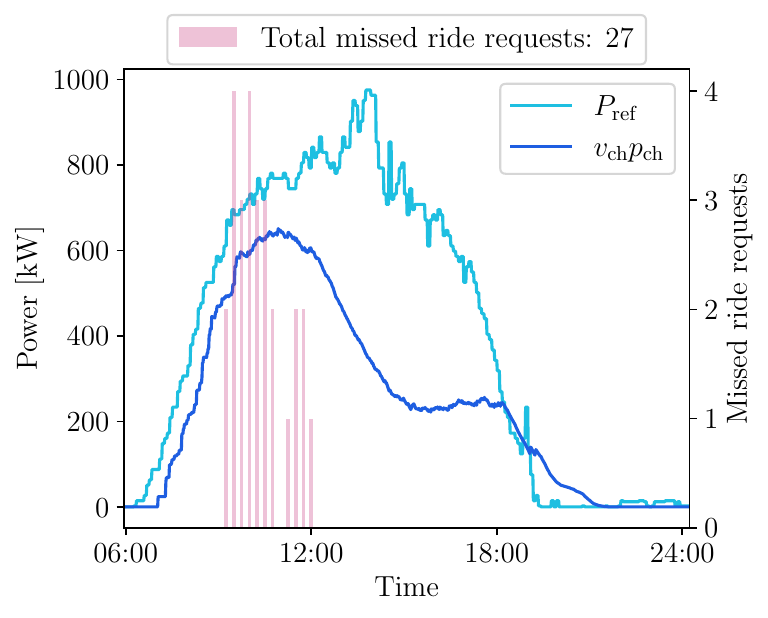}
  \caption{}
  \label{fig:75Su_PrefMissRide_avg}
\end{subfigure}
\centering
\begin{subfigure}{.32\textwidth}
  \centering
  \includegraphics[width=\textwidth]{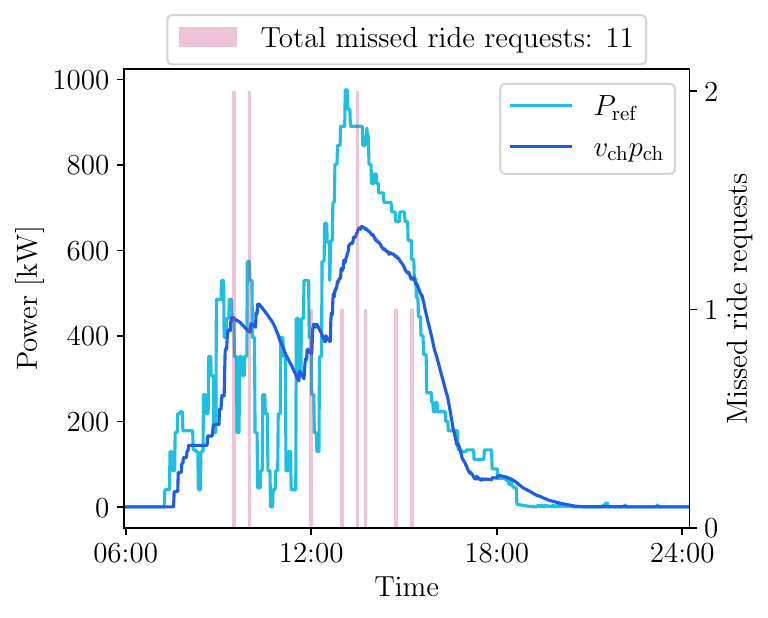}
  \caption{}
  \label{fig:75AM_PrefMissRide_avg}
\end{subfigure}
\centering
\begin{subfigure}{.32\textwidth}
  \centering
  \includegraphics[width=\textwidth]{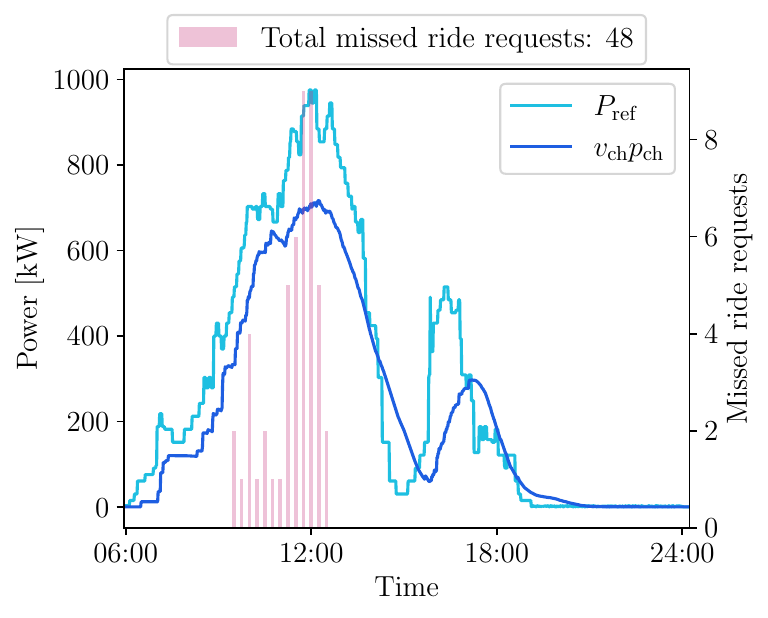}
  \caption{}
  \label{fig:75PM_PrefMissRide_avg}
\end{subfigure}
\centering
\begin{subfigure}{0.32\textwidth}
  \centering
  \includegraphics[width=\textwidth]{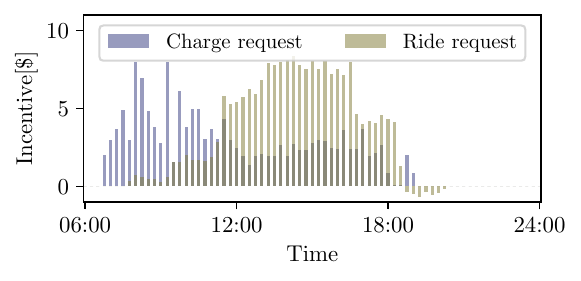}
  \caption{}
  \label{fig:75Su_incAssign_avg}
\end{subfigure}
\centering
\begin{subfigure}{0.32\textwidth}
  \centering
  \includegraphics[width=\textwidth]{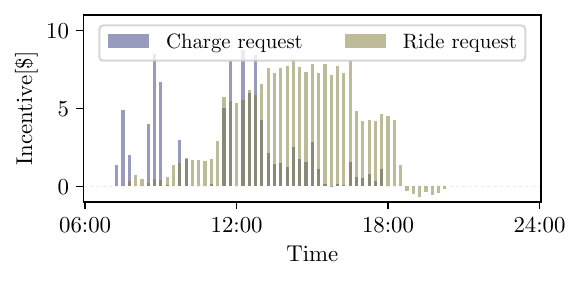}
  \caption{}
  \label{fig:75AM_incAssign_avg}
\end{subfigure}
\centering
\begin{subfigure}{0.32\textwidth}
  \centering
  \includegraphics[width=\textwidth]{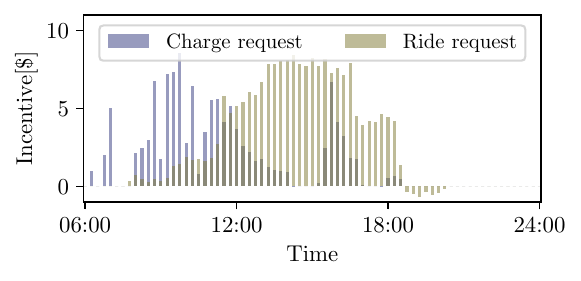}
  \caption{}
  \label{fig:75PM_incAssign_avg}
\end{subfigure}
\caption{Availability of EVs during the day ($1^{\mathrm{st}}$-row), SOC time-evolution ($2^{\mathrm{nd}}$-row), charging profiles together with the total number of missed ride requests ($3^{\mathrm{rd}}$-row) and incentives ($4^{\mathrm{th}}$-row) during the day, given three different weather scenarios: sunny (a)-(d)-(g)-(j), cloudy morning (b)-(e)-(h)-(k), and cloudy afternoon (c)-(f)-(i)-(l). The customers’ willingness to ride-share is set here to 75\%. The initial SOC of each EV is randomly set within $10\%$ and $100\%$ of the battery capacity. The PV-generation ($P_{\mathrm{ref}}$) and charging profiles ($v_{\mathrm{ch}} p_{\mathrm{ch}}$) are obtained summing over all charging stations.}
\label{fig:75}
\vspace{-.3cm}
\end{figure}

\begin{figure}[h!]
\captionsetup{font={small,it},labelfont={bf,sf}}
\captionsetup[sub]{font=footnotesize,labelfont={}}
\centering
\begin{subfigure}{.3325\textwidth}
  \centering
  \includegraphics[width=1.0\textwidth]{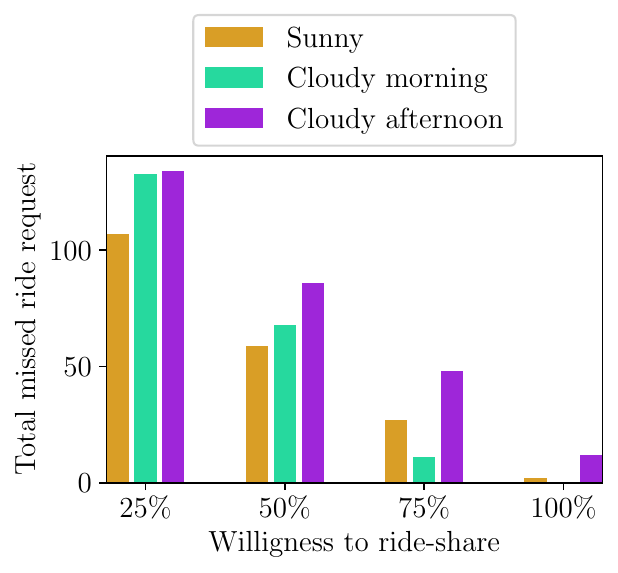}
  \caption{}
  \label{fig:rideSharingwill_missed}
\end{subfigure}
\begin{subfigure}{.610\textwidth}
  \centering
  \includegraphics[width=1.0\textwidth]{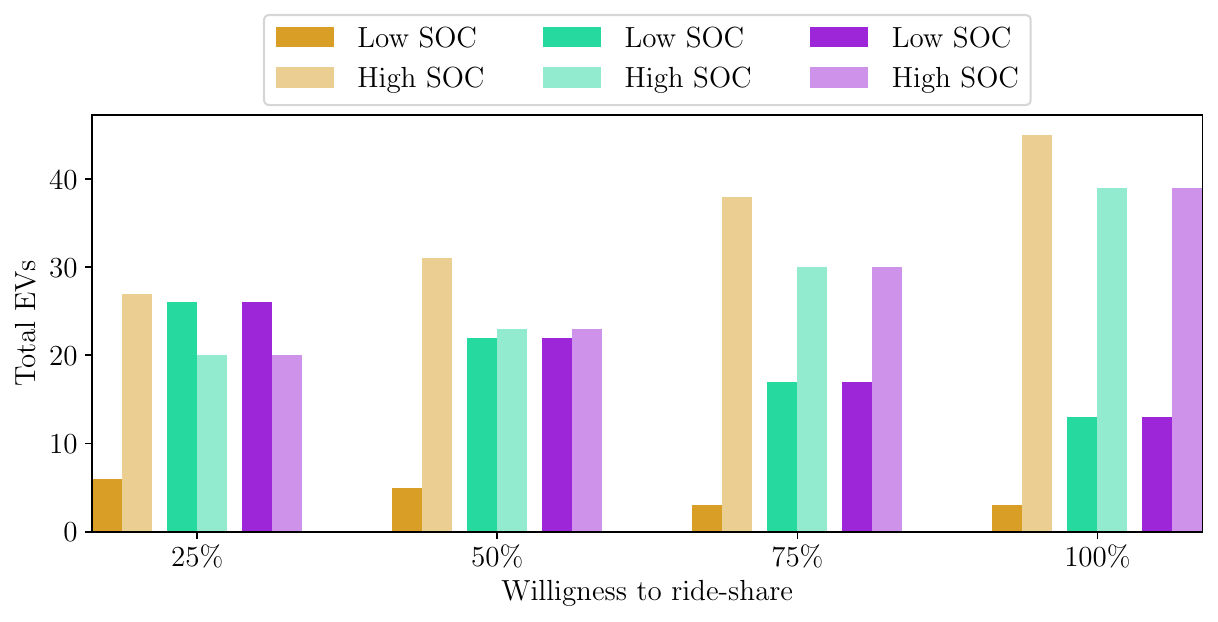}
  \caption{}
  \label{fig:rideSharingwill_EVs}
\end{subfigure}
\caption{Total number of missed ride requests during the day (a) and number of EVs with low/high SOC by the end of the day (b), for given customers' willingness to ride-share and for three different weather scenarios: sunny, cloudy morning, and cloudy afternoon. The initial SOC of each EV is randomly set within $10\%$ and $100\%$ of the battery capacity.}
\label{fig:rideSharing_will}
\vspace{-.3cm}
\end{figure}

We introduce a new parameter, the customers' \textit{willingness to ride-share}, and analyze how it impacts the results in four different scenarios where $100\%$, $75\%$, $50\%$ or $25\%$ of the customers are willing to ride-share. Figure \ref{fig:75} is the counterpart of Figure \ref{fig:Bilevel}, this time taking also ride-sharing ($75\%$) into account. Figure \ref{fig:rideSharingwill_missed} shows how the number of missed ride requests is affected by the customers' willingness to ride-share. For example, assuming that 50$\%$ of the passengers would be willing to share rides on a sunny day, the total number of missed ride requests reduces to 59 and to only 2 in the case where all passengers would agree to share rides. This is a substantial improvement, roughly sixty times less than the case that does not take ride-sharing into consideration. Figure \ref{fig:rideSharingwill_EVs} also reports how many EVs would have a low or high final SOC at the end of the day, depending on the customers' willingness to ride-share. Here, we can see that when the willingness to ride-share increases also the number of EVs with high SOC at the end of the simulation is larger, confirming that an improved QoS can be achieved even if the need for charging is less, as shown in Table \ref{tab:metrics}. 

\begin{figure}[t!]
\captionsetup{font={small,it},labelfont={bf,sf}}
\captionsetup[sub]{font=footnotesize,labelfont={}}
\centering
\begin{subfigure}{.48\textwidth}
  \centering
  \includegraphics[width=\textwidth]{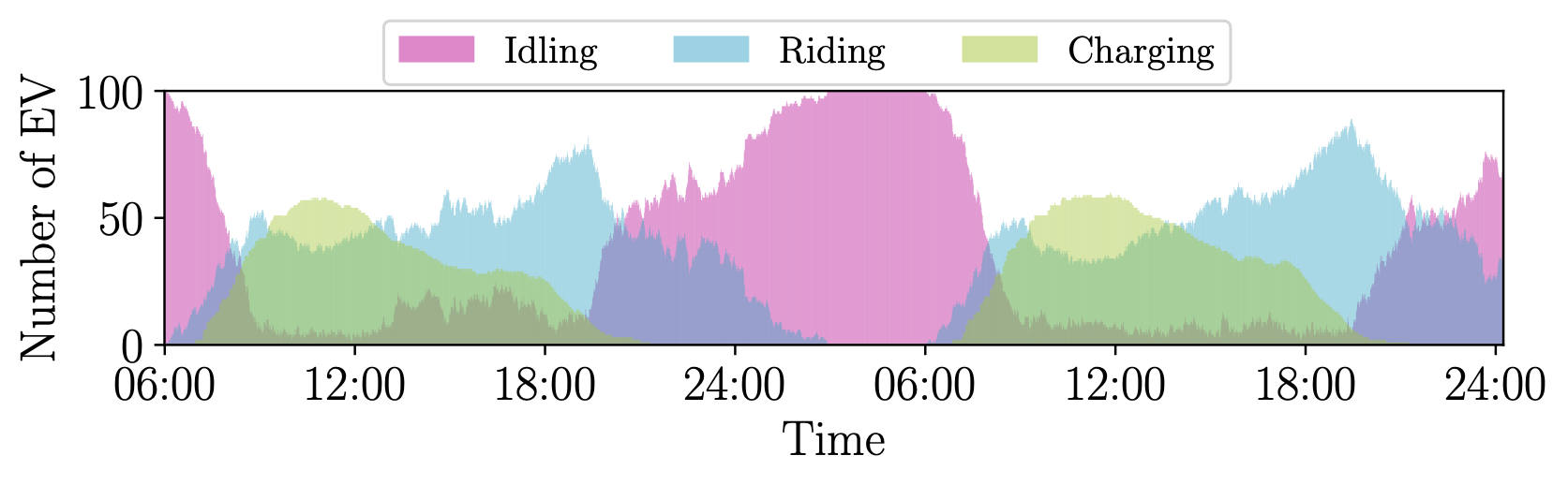}
  \caption{}
  \label{fig:75Su_EV_avg1}
\end{subfigure}
\centering
\begin{subfigure}{.48\textwidth}
  \centering
  \includegraphics[width=\textwidth]{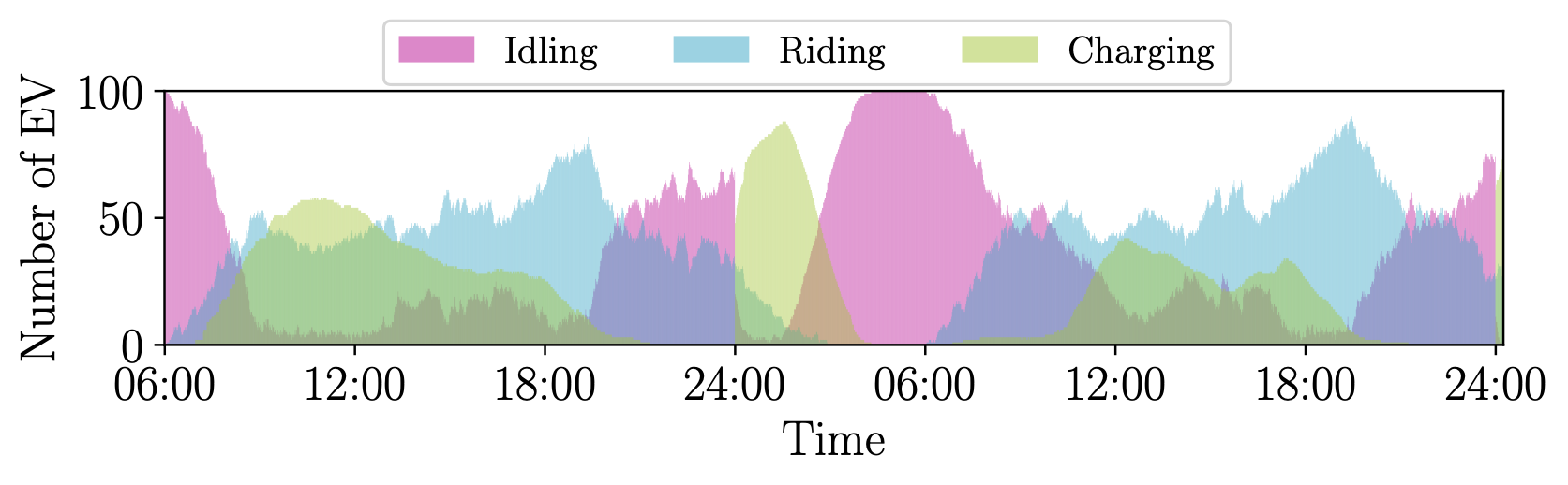}
  \caption{}
  \label{fig:75AM_EV_avg1}
\end{subfigure}
\centering
\begin{subfigure}{.48\textwidth}
  \centering
  \includegraphics[width=\textwidth]{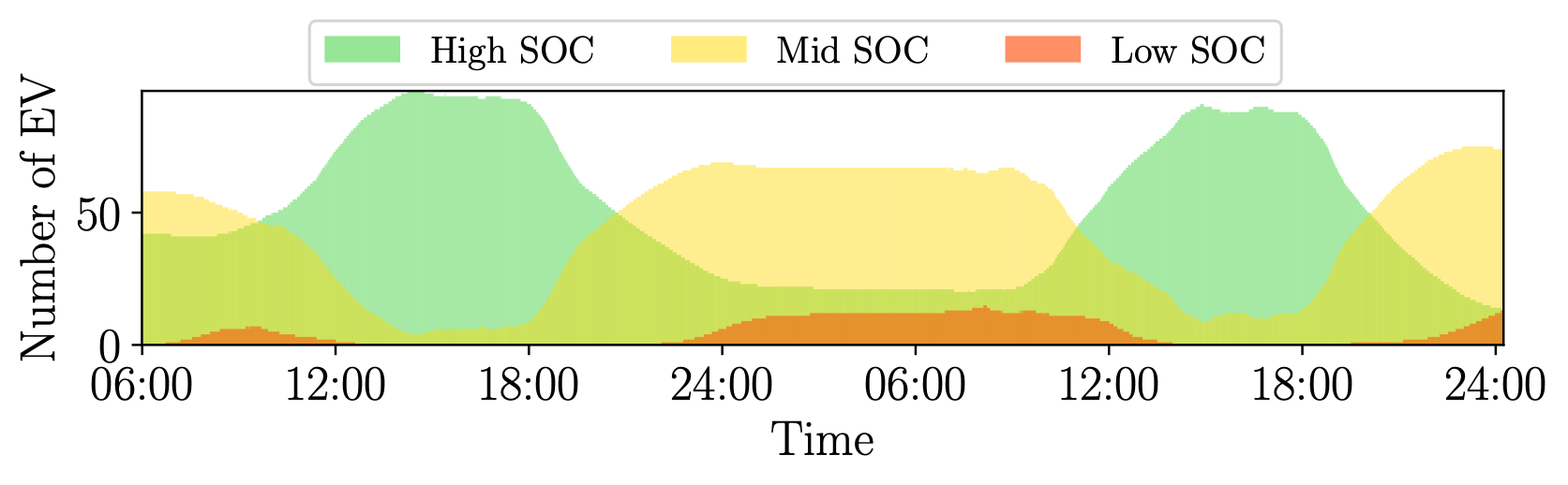}
  \caption{}
  \label{fig:75Su_SOC_avg1}
\end{subfigure}
\centering
\begin{subfigure}{.48\textwidth}
  \centering
  \includegraphics[width=\textwidth]{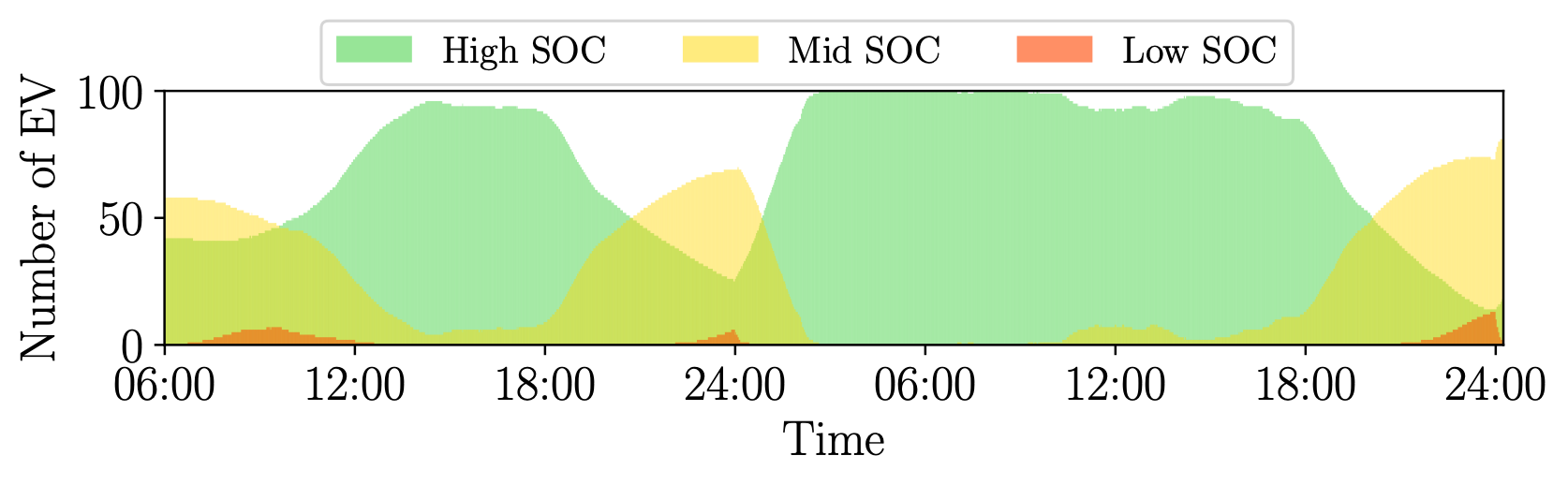}
  \caption{}
  \label{fig:75AM_SOC_avg1}
\end{subfigure}
\centering
\begin{subfigure}{.48\textwidth}
  \centering
  \includegraphics[width=\textwidth]{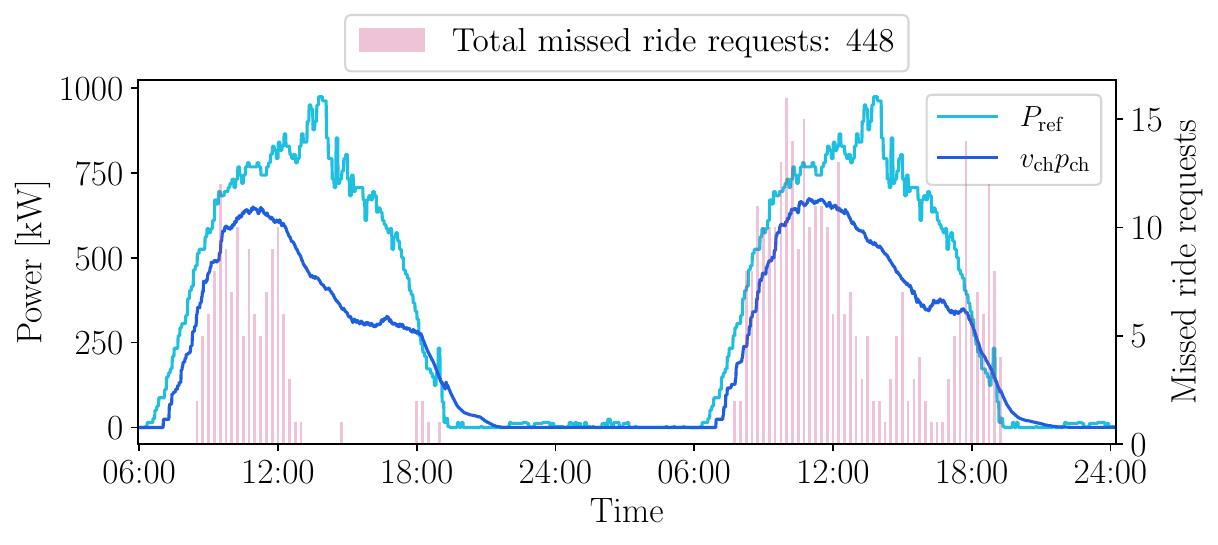}
  \caption{}
  \label{fig:75Su_PrefMissRide_avg1}
\end{subfigure}
\centering
\begin{subfigure}{.48\textwidth}
  \centering
  \includegraphics[width=\textwidth]{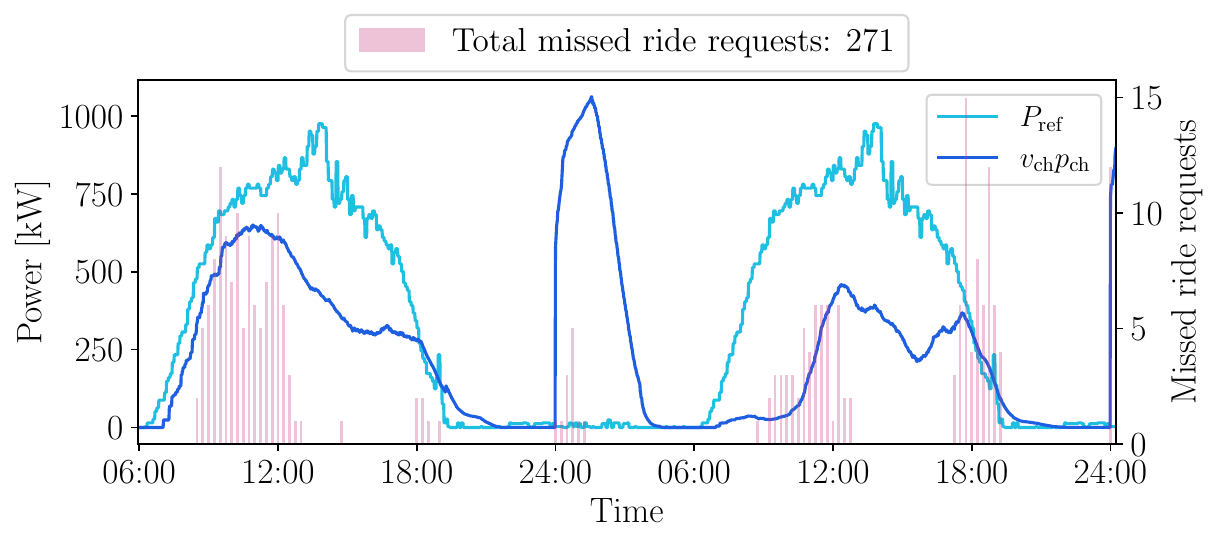}
  \caption{}
  \label{fig:75AM_PrefMissRide_avg1}
\end{subfigure}
\centering
\begin{subfigure}{0.48\textwidth}
  \centering
  \includegraphics[width=\textwidth]{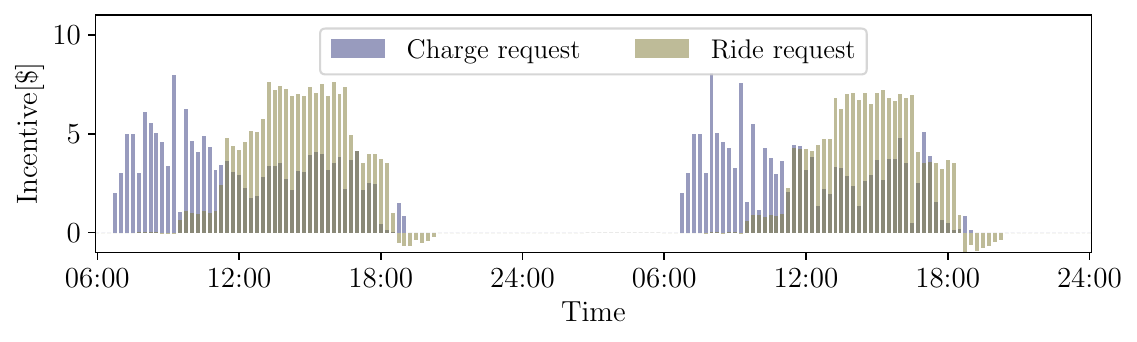}
  \caption{}
  \label{fig:75AM_incAssign_avg1}
\end{subfigure}
\centering
\begin{subfigure}{0.48\textwidth}
  \centering
  \includegraphics[width=\textwidth]{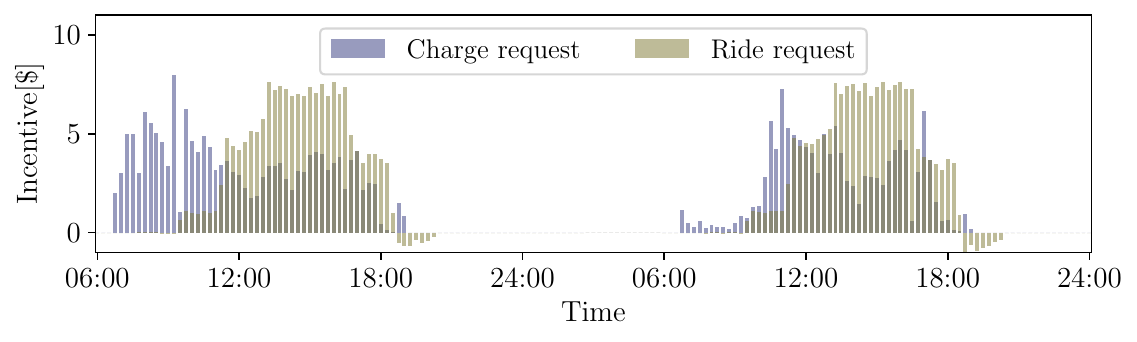}
  \caption{}
  \label{fig:75PM_incAssign_avg1}
\end{subfigure}
\caption{Availability of EVs ($1^{\mathrm{st}}$-row), SOC time-evolution ($2^{\mathrm{nd}}$-row), charging profiles together with the total number of missed ride requests ($3^{\mathrm{rd}}$-row) and incentives ($4^{\mathrm{th}}$-row) during a 42 hour interval with (left) and without (right) additional charging at night. The initial SOC of each EV is randomly set within $10\%$ and $100\%$ of the battery capacity. The PV-generation ($P_{\mathrm{ref}}$) and charging profiles ($v_{\mathrm{ch}} p_{\mathrm{ch}}$) are obtained summing over all charging stations.}
\label{fig:long_run_sharing0}
\vspace{-.3cm}
\end{figure}

\begin{figure}[t!]
\captionsetup{font={small,it},labelfont={bf,sf}}
\captionsetup[sub]{font=footnotesize,labelfont={}}
\centering
\begin{subfigure}{.48\textwidth}
  \centering
  \includegraphics[width=\textwidth]{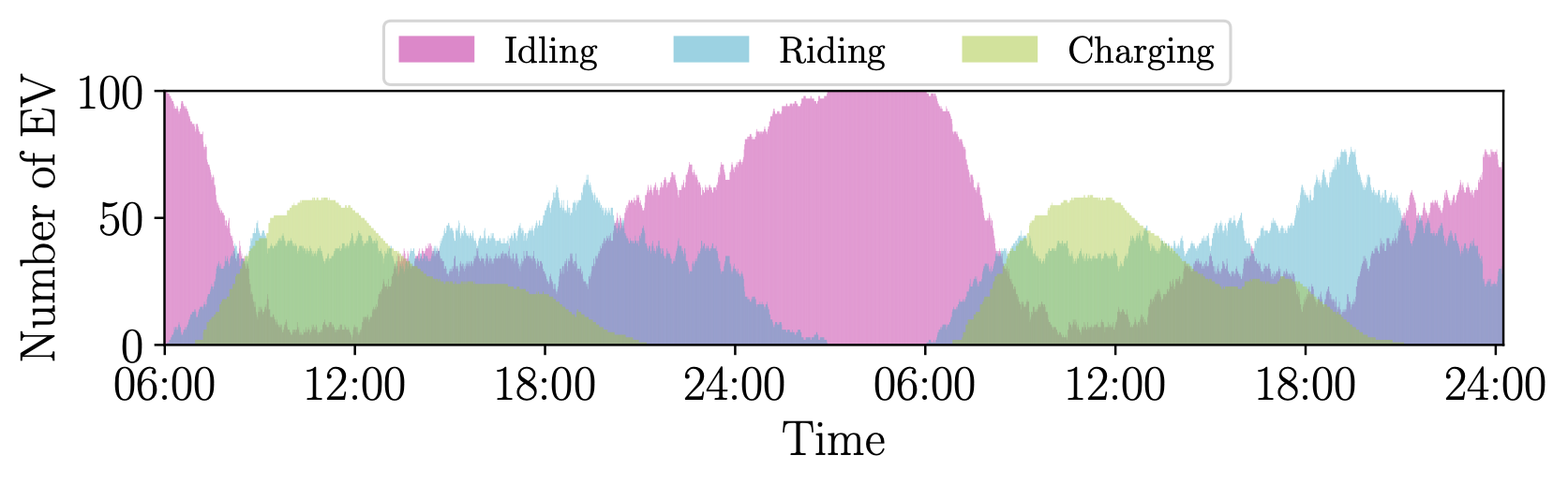}
  \caption{}
  \label{fig:75Su_EV_avg2}
\end{subfigure}
\centering
\begin{subfigure}{.48\textwidth}
  \centering
  \includegraphics[width=\textwidth]{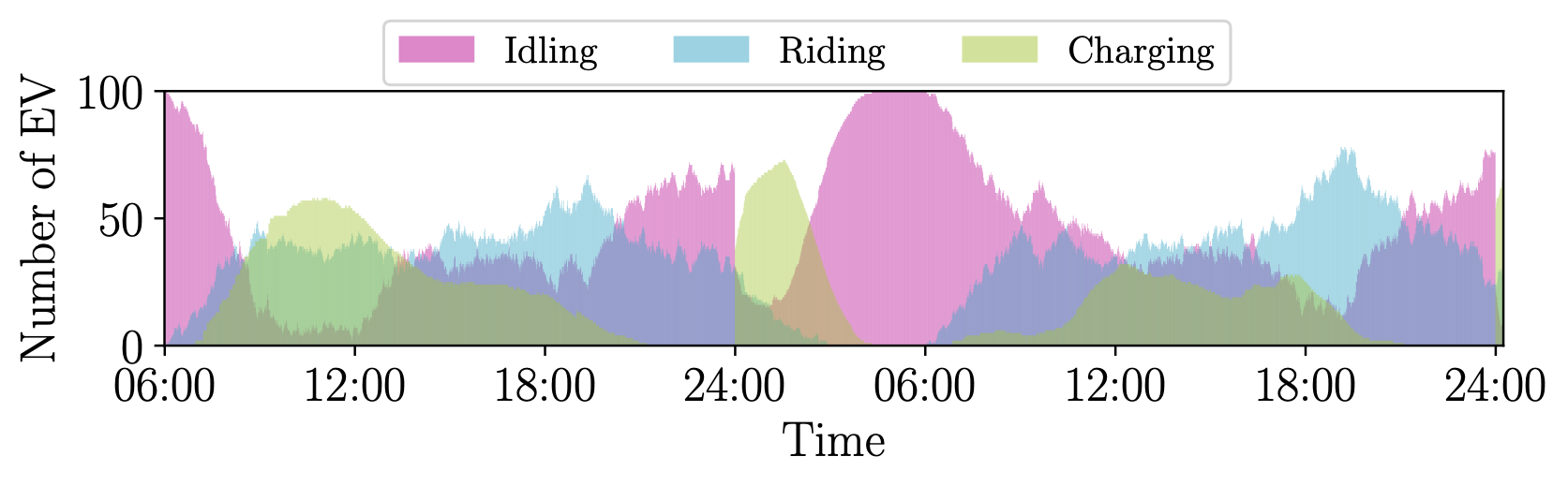}
  \caption{}
  \label{fig:75AM_EV_avg2}
\end{subfigure}
\centering
\begin{subfigure}{.48\textwidth}
  \centering
  \includegraphics[width=\textwidth]{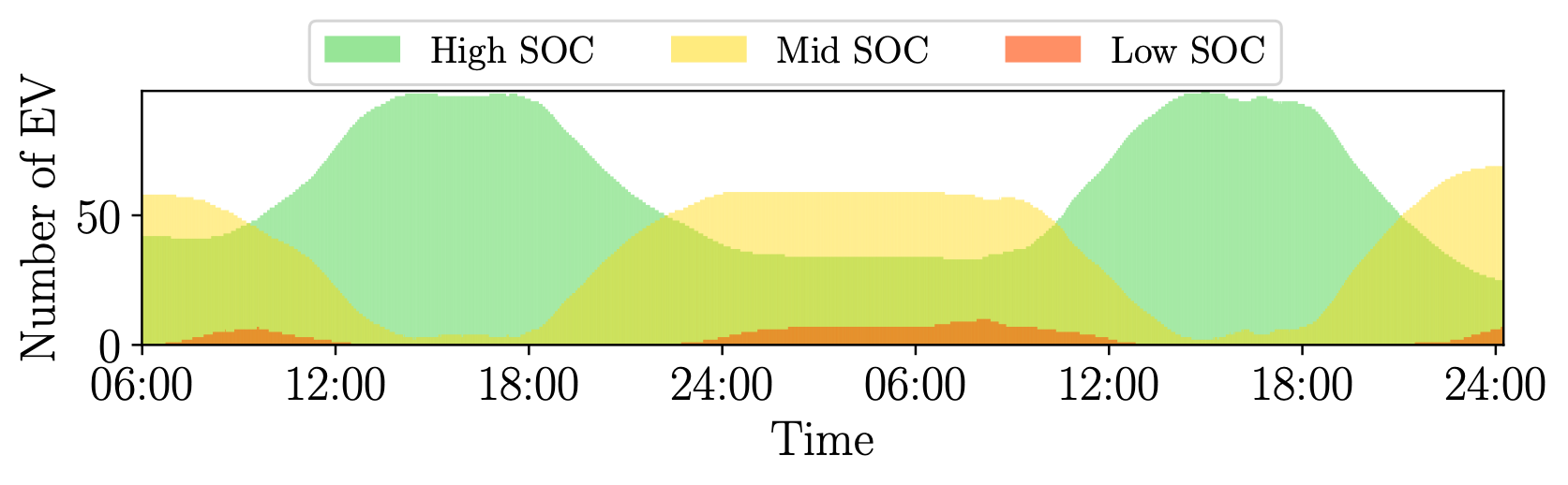}
  \caption{}
  \label{fig:75Su_SOC_avg2}
\end{subfigure}
\centering
\begin{subfigure}{.48\textwidth}
  \centering
  \includegraphics[width=\textwidth]{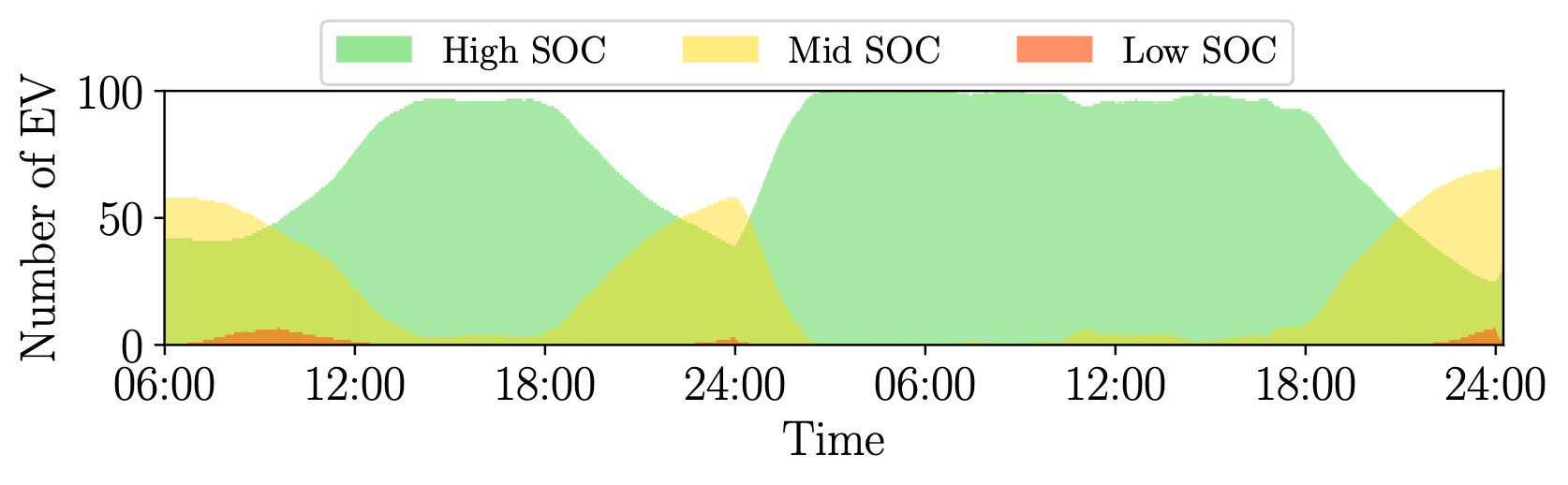}
  \caption{}
  \label{fig:75AM_SOC_avg2}
\end{subfigure}
\centering
\begin{subfigure}{.48\textwidth}
  \centering
  \includegraphics[width=\textwidth]{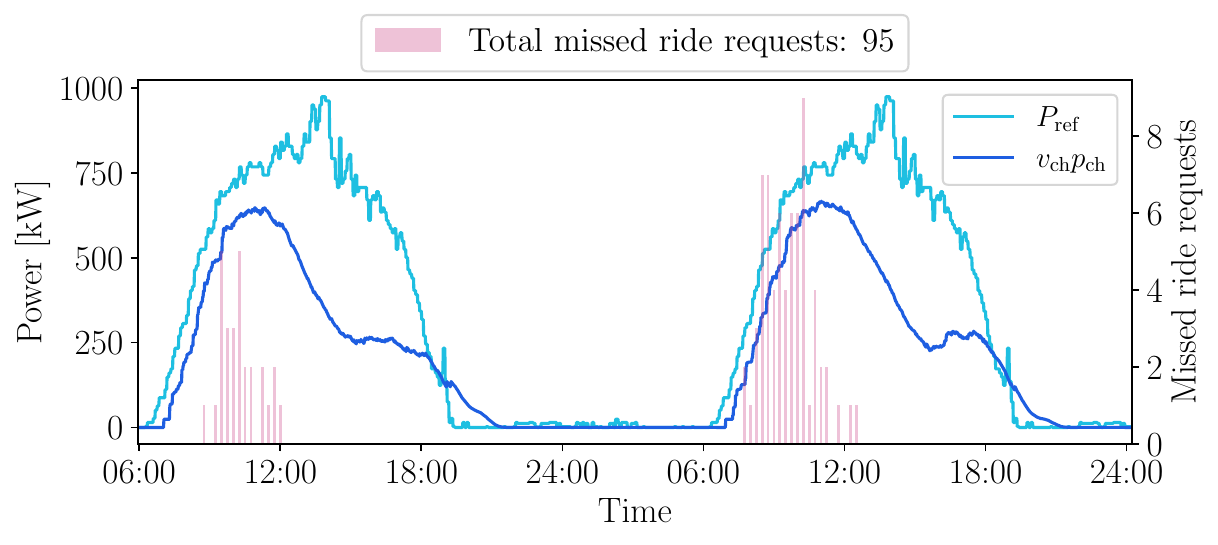}
  \caption{}
  \label{fig:75Su_PrefMissRide_avg2}
\end{subfigure}
\centering
\begin{subfigure}{.48\textwidth}
  \centering
  \includegraphics[width=\textwidth]{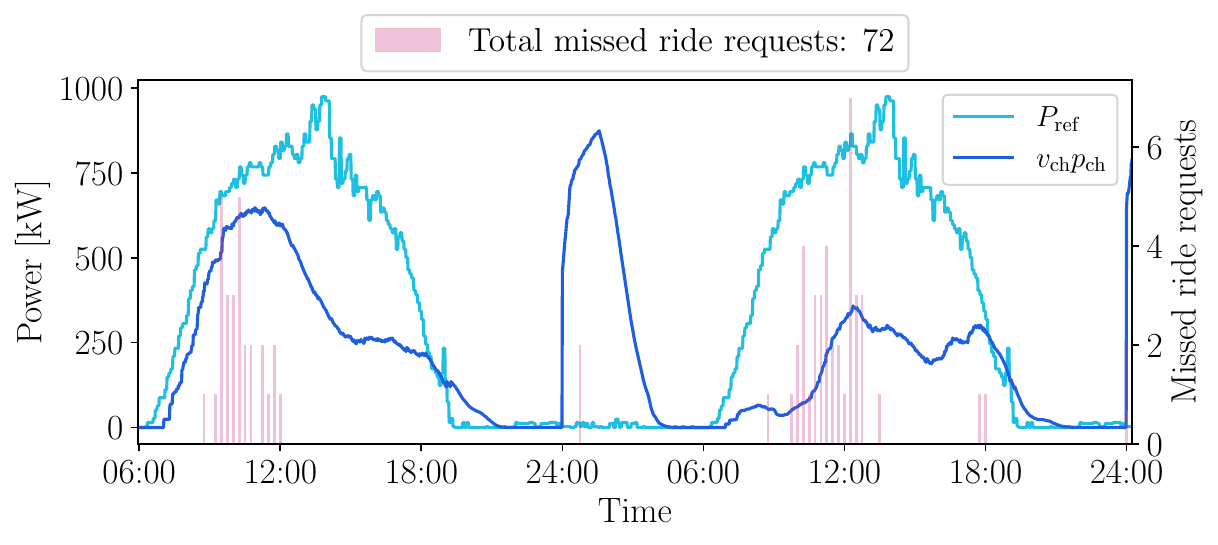}
  \caption{}
  \label{fig:75AM_PrefMissRide_avg2}
\end{subfigure}
\centering
\begin{subfigure}{0.48\textwidth}
  \centering
  \includegraphics[width=\textwidth]{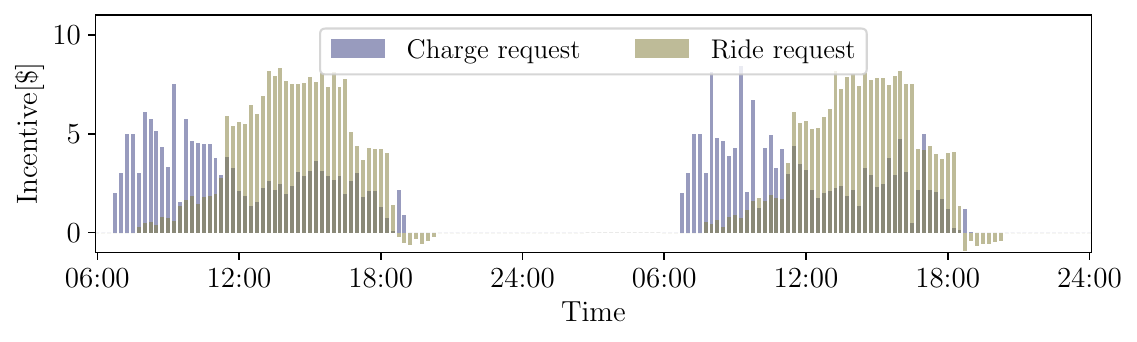}
  \caption{}
  \label{fig:75AM_incAssign_avg2}
\end{subfigure}
\centering
\begin{subfigure}{0.48\textwidth}
  \centering
  \includegraphics[width=\textwidth]{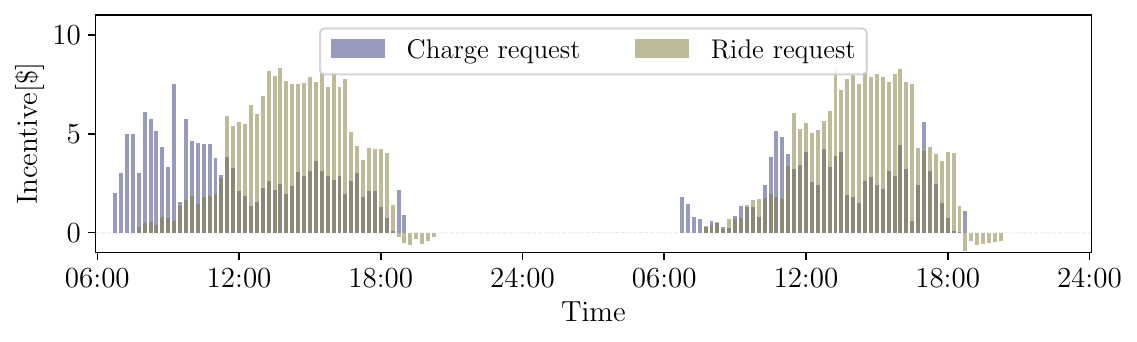}
  \caption{}
  \label{fig:75PM_incAssign_avg2}
\end{subfigure}
\caption{Availability of EVs ($1^{\mathrm{st}}$-row), SOC time-evolution ($2^{\mathrm{nd}}$-row), charging profiles together with the total number of missed ride requests ($3^{\mathrm{rd}}$-row) and incentives ($4^{\mathrm{th}}$-row) during a 42 hour interval with (left) and without (right) additional charging at night. The customers’ willingness to ride-share is set here to 75\%. The initial SOC of each EV is randomly set within $10\%$ and $100\%$ of the battery capacity. The PV-generation ($P_{\mathrm{ref}}$) and charging profiles ($v_{\mathrm{ch}} p_{\mathrm{ch}}$) are obtained summing over all charging stations.}
\label{fig:long_run_sharing75}
\vspace{-.3cm}
\end{figure}


\section*{Discussion}

On-demand ride-sharing helps to reduce traffic congestion and emissions, and especially in densely populated areas it is paving the way for more sustainable mobility. However, it is still facing a significant amount of skepticism due to many heterogeneous factors, ranging from users' preferences to seamless integration into the transportation system\cite{Storch_2021}. In particular, if EVs are adopted as the main way of transportation, the interaction between the ride-service providers and the distribution power system plays a crucial role in determining the successful operation of the EV fleet. 

In this work, we investigate the interaction between a ride-service provider that manages a 100\% EV fleet, and a power utility company that operates renewable energy resources at some charging locations. We show that through the proposed bargaining mechanism we achieved an EV charging schedule that maximizes the use of renewable generation and reduces the potential negative effects of the EV charging on the power infrastructure. Moreover, we are able to preserve the QoS of the \emph{businesses-as-usual} case while improving the impact of the EV charging on the grid. The latter is achieved by scheduling EVs to charge via charge requests issued by the power utility company so that the peak of renewable generation is covered, reducing losses for the power utility company and improving the power grid load balance. The advantages of our method are multifold: exploit renewable energy resources when available, reduce the need for renewable generation storage at a grid level, and charge a large number of EVs before evening rush hours. 

Assuming an increasingly widespread adoption of ride-sharing in the coming years, several implementation strategies have been proposed. To name a few examples representing different perspectives, the design of price mechanisms for ride-sharing\cite{Storch_2021}, a dynamic vehicle-request assignment strategy for autonomous ride-sharing services\cite{JA-SS-AW-EF-DR:17}, a linear vehicle-request assignment within a federated optimization architecture\cite{AS-LM-CG:19}. Other approaches design price mechanisms where the charging facilities set prices to control the resource utilization\cite{Santoyo_2023}, or where the EV charging schedules are influenced by varying energy prices\cite{Mahnoosh_2017}. Here, we propose a novel approach based on a game-theoretic framework that balances the objectives of a ride-service provider, an EV fleet, and a power utility company through a bargaining mechanism. The EV charging strategy influences the charging pattern by controlling a financial incentive sent to the ride-service provider, which in turn solves a linear assignment problem to determine how to best meet both ride and charge requests. The coordination of centralized and decentralized strategies for the EV charging is studied in the context of non-cooperative games in works such as\cite{Ma_2013}, where a large EV population is coupled through a common price signal. However, those approaches do not deal with a ride-service provider trying to serve ride requests and respond to utility needs simultaneously. 

In this work, we show how to integrate the needs of a power utility company, in the form of maximizing the use of renewable generation,  with the users' need to complete trips from an origin to a destination point. The aim is to find an equilibrium between the objectives of different entities. The reason for establishing a charging schedule can be found in the \emph{business-as-usual} case presented in Figure \ref{fig:NoChargeReq}, where no charge requests are submitted and the EVs connect to the power grid whenever their SOC is below a certain threshold. Note that in this case the charging profile is wider and shifted towards the evening hours with respect to Figure \ref{fig:Bilevel}, implying worsened grid loading during an already critical moment of the day, according to the data dashboard of New York ISO for the Manhattan area\cite{NY_ISO_2023}. With our method, \emph{case 1} in the Results section, the charging profile follows the renewable generation profile instead; this contributes to alleviating the so-called ``duck curve'' by promoting charging during high renewable generation periods. We use 63.3\% of the available renewable power to schedule the EV charging (see Figure \ref{fig:Bilevel}) while slightly improving the QoS with respect to the \emph{business-as-usual} case. Additionally, when we introduce the ride-sharing option, as presented in \emph{case 2}, we further improve the QoS while still relying on renewable resources for the EV charging, as presented in Figure \ref{fig:75}-\ref{fig:rideSharing_will} and Table \ref{tab:metrics}. Our findings indicate that ride-sharing is a must in order to fully benefit from an EV fleet powered by renewables. Improved QoS, a larger number of EVs with high SOC, and less traffic congestion are just some of the multiple advantages of ride-sharing.

Note that in this work the power utility company generates only positive incentives since a charge request can only be sent if renewable power is available. The negative incentives (disincentives) with respect to the ride requests seen in Figure \ref{fig:Bilevel}-\ref{fig:75} correspond to the scenario where the users did not submit a bid (tip) and/or some EVs sent a negative incentive to the ride-service provider to indicate that they were far away from the user's drop-off point. 

For completeness, we repeat the simulation for a fleet of regular fossil-fuel vehicles, as shown in Figure \ref{fig:fossil_fuel}. The number of missed ride requests, in this case that does not include the ride-sharing option, reduces to 4. In order to achieve the same result with EVs, a larger fleet with 35 additional EVs would be needed, or alternatively, the customers' willingness to ride-share should be above 90\%. On the other hand, our mobility model is significantly more sustainable since it is entirely powered by renewable energy resources. We emphasize that we can make the EV charging schedule match the renewable generation profile at each charging facility through our bargaining mechanism, as shown in  Figure \ref{fig:Bilevel_power4sation}. Moreover, we note that our results may further improve in a scenario where fast-charging is adopted. In this work, we take a conservative approach and consider level 2 chargers only, but if we consider level 3 EV chargers, also called DC fast chargers\cite{EV_level3}, we could achieve a considerable improvement in the QoS. For example, if we assume a charging rate of 1.15 kWh per minute then the time needed to recharge a depleted battery would reduce to 45 minutes, and the performance metrics for \emph{case 1} on a sunny day would improve, resulting in a QoS of 99.8\% and a PL of 37.4\%.

To provide a more comprehensive set of results, we examine how our method performs over a longer simulation period. Figure \ref{fig:long_run_sharing0} and \ref{fig:long_run_sharing75} show the results over 42 hours for \textit{case 1} and \textit{case 2}, respectively. Moreover, we consider an additional case in the right column, where we assume that the EVs are allowed to charge at night, during off-peak hours, by connecting to the grid regardless of where they are located. For comparison, the total number of ride requests is now 5156, out of which 2676 were received on the second day. As a benchmark metric, the total number of missed ride request for the \emph{business-as-usual} case is 450. We notice a significant improvement in the QoS when allowing grid charging at night, in order to restore higher SOC levels while EVs are less busy driving.

We acknowledge that our formulation comes with limitations. First, we consider charge requests as an input that depends on the surplus of renewable generation in areas where the charging facilities are located. With coordinated efforts between the ride-service provider and the power utility company, it would be feasible to displace the EV charging peak to a time period with minimal arrival of ride requests, mitigating the adverse effects of charging on the QoS, as currently can be seen in Figure \ref{fig:Bilevel}. However, we note that the proposed approach can be extended to consider the case where the power utility company incentivizes the EVs to provide other services, such as ramping services and load balancing; this is performed by replacing the renewable generation with a different power profile in the proposed method. Second, the performance of our algorithm relies on the tuning of multiple parameters that affect the outcome for the ride-service provider, by improving the QoS, or for the power utility company, by increasing the consumption of renewable generation; then, the balance between the objectives of both companies is not a trivial task. Our algorithm setup depends on external variables such as the user bid, energy prices, initial SOC of EVs, and arrival time of ride and charge requests, thus, the results can differ considerably from one case to another. In particular, we recognize the importance of setting the value for the maximum bid allowed to users, denoted by $h_{\max}$. If users can bid high enough to request a trip, ride requests will be prioritized over charge requests increasing the PL for the power utility company. Otherwise, if users do not offer bids at all, and the financial incentives associated with the charge requests are large enough, the QoS can deteriorate due to the preference of the ride-service provider to prioritize charging. 

As a final remark, we note that the proposed assignment strategy does not plan rides and charging profiles ahead based on predictions of demand and power generation. In line with other existing works such as \cite{Rossi_2020, AS-LM-CG:19, Dafernos_1969, JA-SS-AW-EF-DR:17}, we seek an assignment problem that involves a computationally light solution and can be solved at the minute or sub-minute time scale. Extending the framework to a rolling horizon optimization setup (as in, for example, model predictive control frameworks) to include the planning of rides and charging profiles ahead would significantly increase the computational complexity of our method; this is mainly because the relaxation of the binary variables may not be exact. It is also important to note that, when planning ahead, one should consider errors in the predictions of demand and power generation. Exploring robust or stochastic problems  for multi-period problem formulations along with computationally affordable solution approaches goes beyond the scope of the present paper and will be the subject of future research endeavors.

\section*{Methods}

In this section, we outline the main mathematical framework utilized to develop the proposed bargaining mechanism, and we explain the main implementation. The proposed mechanism involves three entities: a ride-service provider, a fleet of EVs (either with a driver or autonomous), and a power utility company; they are shown in Figure \ref{fig:cartoonA}. These three entities interact each time a new assignment has to be made.

In the proposed framework, the power utility company aims to maximize the use of renewable generation at specific portions of the grid, where charging facilities powered by renewable resources are located. The ride-service provider manages the EV fleet to serve ride requests from customers and respond to the power utility company requests while minimizing its overall operational costs. The goal of the EV drivers is to get ride requests assigned to their EVs.

The proposed mechanism is repeated at given intervals (e.g., $1, 5, 10, ...$ minutes, depending on specific settings) to assign available vehicles to new ride and charge requests. Accordingly, time is discretized as $t \in \{0, 1, 2, \dots \}$ (normalized to integer units). At each given time  $t$, the scheme involves three main steps, illustrated in Figure \ref{fig:cartoonA}: 

\noindent \emph{Step~1}. Customers send \emph{ride requests} to the ride-service provider, accompanied by a possible additional tip they are willing to offer in order to compete against other customers that are also requesting rides; in parallel, the power utility company sends to the ride-service provider a number of \emph{charge requests}.  

\noindent \emph{Step~2}. The EVs, the ride-service provider, and the power utility company start the bargaining procedure, mathematically explained in  Algorithm~1 (outlined shortly). This step involves an iterative procedure where the ride-service provider computes potential EV-request assignments, and communicates them to the EVs and power utility company, which in turn provide new incentives to possibly influence the assignment; the interaction then repeats with the ride-service provider re-computing the potential EV-request assignments.  

\noindent \emph{Step~3}. Once Step~2 ends, the ride-service provider issues the final assignments. 

In the following, we describe the mathematical problem formulation and pertinent algorithms associated with Steps~1--3. 


\subsection*{A New Vehicle Assignment Method Based on Renewable and Ride Incentives}

We model the transportation network topology as exemplified in Figure \ref{fig:cartoonB}. Movements of EVs between geographical areas are described by an undirected graph; the nodes $\cN = \{1, 2, \ldots, n\}$ of the graph represent $n$ geographical areas (neighborhoods, groups of city blocks, or towns depending on the geographical granularity); two areas are connected through an edge if they can be reached within a given traveling time. Assume that the ride-service provider receives $p$ requests, as illustrated in Figure \ref{fig:cartoonA}, and these are indexed in by set $\cR = \{1, 2, \ldots, p\}$; each ride request includes an origin and destination. The set  $\cC = \{1,2, \ldots, q\}$ represents the charge requests issued by the utility company; each request is associated with a charging facility (located in one of the areas $\cN$). We define $\cE = \cR \cup \cC$ to be the set of all requests (ride and charge), and we let the set of EVs be $\cV = \{1, 2, \ldots, m\}$, where $m$ is the numbers of EVs. At every time slot $t$, the sets $\cV, \cR, \cC$ are updated based on the current availability of vehicles and the most updated ride and charge requests. For notational readability, in what follows we will drop the dependence of these sets on the variable $t$.

In the optimization problems described in the following, we have three sets of optimization variables: (i) the variables $x_{ij} \in \{ 0,1 \}$ are used to describe whether the EV $i$ is assigned or not to the ride or charge request $j$ (with $i \in \cV$ and $j \in \cE$); these are decision variables of the ride-service provider. (ii) The variables $\{y_{ij}, i\in \cV, j \in \cC\}$ are decision variables of the power utility company, and represent financial incentives offered to the ride-service provider in order to incentivize the assignment of EVs to issued charge requests. (iii) The set of variables $\{y_{ij}, j \in \cR\}$ represents financial incentives computed by each EV $i$ and sent to the ride-service provider in order to obtain the ride request $j$. In other words, any available EV $i$ has the task to determine the size of the incentive $\{y_{ij}, j \in \cR\}$ based on the user's bid for request $j$, and on the cost of the potential trip, from the current position of EV $i$ to the users' drop-off location. The complete list of parameters and variables is provided in Table \ref{tab:Nomenclature}. 

With these three groups of variables, the optimization problems associated with the ride-service provider, the EVs, and the power utility company are explained next. We will first outline the three optimization problems, and then explain how these three problems are integral parts of the proposed bargaining mechanism in Step~2. 

\vspace{.2cm}

\noindent \emph{\textbf{Linear Assignment at the Ride-Service Provider.}} We begin by formalizing the optimization problem that is solved at the ride-service provider to assign requests to available vehicles. We recall that we use binary variables $x_{ij} \in \{0, 1\}$ to describe whether EV $i$ is assigned or not to the request $j$; i.e., if $x_{ij}=1$, then EV $i$ is assigned to serve the ride/charge request $j$, otherwise, $x_{ij}=0$. These are referred to as \emph{assignment variables}. Let $c_{ij}$, for $i \in \cV$ and $j \in \cR$, be the fixed cost (in USD, EUR, etc.) for the ride-service provider to attend the ride request $j$ using the EV $i$. On the other hand, $d_{ij}$ is the cost for the ride-service provider to attend the charge request $j$ via EV $i$ in a given charging facility $s$. The variables $y_{ij}$ can be seen as a discount price (respectively, a price increase) to induce the EV $i$ to be assigned to (respectively, not to be assigned to) the ride or charge request $j$.

When the incentives $\by:=\{y_{ij}, i \in \cV, j\in\cE\}$ are given, the operational cost of the ride-service provider can be minimized through the following linear assignment problem: 
\begin{subequations} \label{eq:RHC_pro}
\begin{align}
\textrm{RSP($\by$)}~~~~\min_{\{x_{ij} \in \{0,1\}, i \in \cV, j \in \cE\}} ~~ & \underbrace{\sum_{i \in \cV} \left( \sum_{j \in \cR } (c_{ij} - y_{ij}) x_{ij} + \sum_{j \in \cC } (d_{ij} - y_{ij}) x_{ij} \right)}_{:= f(\bx, \by)},  \label{eq:RHC_pro_cost} \\
\text{s.t.} ~~~~~& \sum_{i \in \cV} x_{ij} = 1 \quad \forall j \in \cE,  \label{eq:RHC_pro_C1} \\
& \sum_{j \in \cE} x_{ij} \leq 1 \quad \forall i \in \cV. \label{eq:RHC_pro_C2}
\end{align}
\end{subequations}
where the label ``RSP($\by$)'' emphasizes that this problem is solved by the ride-service provider once the incentives $\{y_{ij}, i \in \cV, j\in\cE\}$ are given (they are inputs to the problem). Constraints \eqref{eq:RHC_pro_C1} guarantee that each ride/charge request is assigned to one EV, and constraints \eqref{eq:RHC_pro_C2} ensure that each EV can be assigned to at most one request at time $t$. The solution to the optimization problem~\eqref{eq:RHC_pro} minimizes the operational costs of the fleet by optimally assigning the available EVs to ride and charge requests. As in existing linear assignment problems, for the ride requests we consider the cost $c_{ij}$ as the cost of the shortest path between the position of EV $i$ and the pick-up point corresponding to the ride request $j$; similar arguments apply to $d_{ij}$ for the charge request $j$ corresponding to a charging facility.

We note that~\eqref{eq:RHC_pro} is a mixed-integer linear program (MILP), which may become computationally burdensome with the increasing of number of vehicles, ride requests, and charge requests. However, we will show later in the paper that one can take advantage of the totally unimodular constraint matrix property of the linear assignment problem \eqref{eq:RHC_pro}, and show that a continuous relaxation (i.e., substitute $x_{ij} \in \{0, 1\}$ with $x_{ij} \in [0, 1]$) is exact \cite{AS-LM-CG:19}. This, in turn, allows one to leverage standard solvers for linear programs to find an optimal assignment.

\vspace{.2cm}

\noindent \emph{\textbf{Charge Incentives at the Power Utility Company.}} We consider an optimization problem solved by the utility company to minimize the economic loss due to the unused renewable generation at specific charging facilities. To this end, let $\cS$ be the set of charging facilities (located within the neighborhoods or areas $\cN$). We let $\{y_{i{j}}\}_{i\in\cV,  j \in \cC} \mapsto U_s(x_{i {j}}, y_{i{j}}, \theta_s)$ be a function modeling a financial or operational cost incurred by the power utility company at the charging facility $s\in \cS$ for not using power from renewable sources of energy; this function is parametrized by the assignments $\{x_{ij}\}_{i\in\cV,  j \in \cC}$ and by additional parameters $\theta_s \in \mathbb{R}^\theta$ that are of interest to the utility company (examples are given shortly in the section Experimental Setup). Moreover, we let $\{y_{i{j}}\}_{i\in\cV,  j \in \cC} \mapsto \rho_s (x_{i {j}}, y_{i{j}})$ be a function that keeps track of the total incentives assigned at a charging facility $s$. With this notation, and for a given (potential) vehicle-request assignment $\bx_\cC := \{x_{ij}, \,i\in\cV,  j \in \cC\}$, the power utility company solves the following problem to compute the financial incentives associated with its charge requests:
\begin{subequations} 
\label{eq:PUC_pro} 
\begin{align} 
\textrm{PUC($\bx_\cC$)}~~~~\min_{\{y_{ij} \in \cY_\cC: \,i\in\cV,  j \in \cC\}} ~~ &  \sum_{s \in \cS} U_s (y_{i{j}}; \bx_\cC, \theta_s ), \\
\text{s.t.} ~~~~~& b_{\min,s} \leq \rho_s(y_{ij};\bx_\cC) \leq b_{\max,s}, \; \forall s \in \cS.
\end{align}
\end{subequations}
where $\cY_\cC$ is a convex set. The label ``\textrm{PUC($\bx_\cC$)}'' once again stresses that this is a problem solved by the power utility company, for a given assignment $\bx_\cC$. Throughout this work, we assume that the function $U_s$ is convex for any fixed $x_{ij}$ and $\theta_s$, while the function $\rho_s$ is linear or affine; consequently,~\eqref{eq:PUC_pro} is a convex problem.  

\vspace{.2cm}

\noindent \emph{\textbf{Ride Incentives at the EV Fleet.}} We include in our framework an optimization problem associated with each EV, utilized by the EV owner to incentivize the ride-service provider to assign their preferred rides. To this end, let $\bx_\cR := \{x_{ij}, \,i\in\cV,  j \in \cR\}$ denote the ride-requests assignment, and 
$\{y_{ij}\}_{j \in \cR} \mapsto D_{ij}(y_{i{j}}; \bx_\cR, w_{ij})$ be a function modeling a financial cost incurred by the $i$-th EV when serving the ride assignment described by $\bx_\cR$, where $w_{ij} \in \real$ parametrizes the cost. Then, each EV $i$ solves the following ``best bid problem:'' 
\begin{equation} \label{eq:EVi_pro}
\textrm{EV-$i(\bx_\cR)$}~~~~\min_{\{y_{ij} \in \cY_\cR: \, j \in \cR\}} ~~ \sum_{j \in \cR } D_{ij}(y_{ij}; \bx_\cR, w_{ij}).
\end{equation}
Here, $\cY_\cR$ describes a convex set of operational constraints, and $D_{ij}$ is assumed to be a convex function for each fixed $x_{ij}, w_{ij}$. More precisely, in this context, $y_{ij}$ represents the (dis)incentive that the ride-service provider will receive from the EV owner if the EV $i$ is assigned to the ride request $j$. 

The cost in~\eqref{eq:EVi_pro} can also be decoupled as $D_{ij}(y_{ij}, w_{ij}) = D^\prime_{ij}(y_{ij}, w_{ij}) + D^{\prime\prime}_{ij}(y_{ij}, b_{ij})$, where $D^\prime_{ij}(y_{ij}, w_{ij})$ is a term related to the potential financial gain of the driver when taking a ride, and $D^{\prime\prime}_{ij}(y_{ij}, b_{ij})$ is related to the SOC level or the (un)willingness of the driver to take a ride request $j$. Of course, these costs can be engineered based on specific implementations of the algorithms, market setups, and models for the preferences of the drivers.  As an example for the first term, the parameter $w_{ij} \in \mathbb{R}$ may represent a fixed (dis)incentive defined as $w_{ij} = h_{j} - \alpha_{ij} a_{ij}$, where $h_{j}$ is the bid (or tip) offered by the customer that sent the ride request $j$ (satisfying $ 0 \leq h_{j} \leq h_{\mathrm{max}}$), $a_{ij}$ is the cost of reaching the drop-off location of the ride request $j$ starting from the current position of EV $i$, and $\alpha_{ij} \in [0, 1]$ is a scaling factor decided by the EV owner; the cost can be set to, as an example, $D^\prime_{ij}(y_{ij}, w_{ij}) = (y_{ij} - w_{ij})^2$.  As an additional example, in the ride-sharing context, $w_{ij}$ can be given by $w_{ij} = h_{j} - \alpha_{ij} a_{ij} + \beta b_i$, where $\beta \in [0,1]$, and $b_i$ represents an additional incentive that depends on the number of currently available seats on EV $i$. As an example for the second term, one may consider the cost $D^{\prime\prime}_{ij}(y_{ij}, b_{ij}) = y_{ij} b_{ij}$, where the parameter $b_{ij}$ can be inversely proportional to the SOC; moreover, the term $b_{ij}$ may take high values for rides that lead to a low SOC upon completion of the trip.
We would like to point out that~\eqref{eq:EVi_pro} is no longer necessary when considering a fleet of autonomous vehicles.

\vspace{.2cm}

\noindent \emph{\textbf{How do Power Utility Company, Ride-Service Provider, and EVs interact?}} The optimization problems~\eqref{eq:RHC_pro}--\eqref{eq:EVi_pro} are utilized to compute potential assignments (for given incentives) and incentives (for given potential assignments). We consider a game-theoretic approach where potential assignments and incentives are sequentially updated until they converge; we will comment later in the paper on the properties of the assignments and incentives at convergence (and their alignment with the concept of Nash equilibrium). Considering an iterative approach, where $k \in \mathbb{N} \cup \{0\}$ is the iteration index, let $\by^{(k)}$ and $\bx^{(k)}$ the assignments and incentives at iteration $k$; then, our mechanism involves updates of the form 
$\textrm{RSP}(\by^{(0)}) \rightarrow \bx^{(1)} \rightarrow  \textrm{PUC}(\bx_\cC^{(1)}),\textrm{EV-}i(\bx_\cR^{(1)})  \rightarrow \by^{(1)} \rightarrow \textrm{RSP}(\by^{(1)}) \rightarrow \bx^{(2)} \rightarrow  \textrm{PUC}(\bx_\cC^{(2)}),\textrm{EV-}i(\bx_\cR^{(2)})  \rightarrow \by^{(2)} \rightarrow \ldots$ until convergence; the assignments at convergence are then dispatched to the EVs. The notation $\textrm{RSP}(\by^{(k)})$ means that the problem~\eqref{eq:RHC_pro} is solved by the ride-service provider to issue a new potential assignment $\bx^{(k+1)}$, based on the current incentives $\by^{(k)}$ received from the power utility company and the EVs. Similarly, $\textrm{PUC}(\bx_\cC^{(k)})$ means that the power utility company computes new incentives $\by_\cC^{(k+1)}$ by solving~\eqref{eq:PUC_pro} based on the current potential assignment $\bx_\cC^{(k)}$, and each EV updates its own incentives by solving~\eqref{eq:EVi_pro}. At each round, the ride-service provider sends to the power utility company and EVs new potential assignments and receives from them new incentives. This mechanism, which is in the form of a Gauss-Seidel method\cite{facchinei_2010_NE-VI} and is repeated at every time slot $t$, is tabulated as Algorithm~\ref{algo:Gauss-Seidel}.

\begin{algorithm}[h!]
    \caption{Gauss-Seidel Best Response-based Algorithm}\label{algo:Gauss-Seidel}
    \begin{algorithmic}
        \State \textbf{\# Initialization}
        
        \State \textbf{[I1]} Ride-service provider receives ride requests $\cR$ and charge requests $\cC$. 

        \State \textbf{[I2]} Ride-service provider chooses a feasible initial assignment  $\bx^{(0)}$ and sends it to the power utility company and EVs.

        \State \textbf{\# Gauss-Seidel mechanism}
        
            \For{$k =1, 2, \dots$ until termination criterion is  satisfied}

                \State \textbf{[S1-a]} Power utility company updates incentives as:
                $$\by_{\cC}^{(k)} \in \underset{\by}{\arg \min} \, \left\{ \sum_{s \in \cS} U_s\left(\by; \bx_\cC^{(k-1)}, \theta_s \right) \, | \, {b}_{\min,s} \leq \rho_s(\by; \bx_\cC^{(k-1)}) \leq {b}_{\max,s} ~ \forall s \in \cS, \, \by \in \cY_\cC  \right\}.$$ 

                \State \textbf{[S1-b]} Power utility company sends $\by_{\cC}^{(k)}$ to ride-service provider.

                \State \textbf{[S2-a]} Each EV  updates incentives as:
                $$
                \by_\cR^{(k)} \in \arg \min_{\{y_{ij} \in \cY_\cR , \, j \in \cR\}} ~~  \sum_{j \in \cR } D_{ij} \left(y_{ij}; \bx_\cR^{(k-1)}, w_{ij} \right).
                $$
                
                \State \textbf{[S2-b]} Each EVs send $\by_{\cR}^{(k)}$ to ride-service provider.

                \State \textbf{[S3-a]} Ride-service provider updates the potential assignment as:  

                $$\bx^{(k)} \in \underset{\bx}{\arg \min} \, \left\{ f(\bx; \by^{(k)}) \, | \, \bx \in \cX \right\}, $$

                \State ~~~~~~~~~~~ where $\cX = \left\{ \bX \in \bbR^{m \times p+q} \, : \, x_{ij} \in [0,1], \, \text{and} \, \sum_{i \in \cV'} x_{ij} = 1, \, \forall j \in \cE', \sum_{j \in \cE'} x_{ij} = 1, \, \forall i \in \cV' \right\}$.

                \State \textbf{[S3-b]} Ride-service provider sends $\bx_{\cC}^{(k)}$ to utility company and $\bx_{\cR}^{(k)}$ to EVs.
            \EndFor
            
        \State \textbf{\# EV dispatch}
        \State \textbf{[D1]} Ride-service provider dispatches assignments to EVs.
    \end{algorithmic}
\end{algorithm}

Note that in step [S3-a] the binary variables $x_{ij} \in \{0,1\}$ have been relaxed to $x_{ij} \in [0,1]$; accordingly, the relaxed assignment problem in step [S3-a] is a continuous convex linear problem, which can be solved efficiently. Moreover, in the section Theoretical Foundations, we will show that the relaxation is exact. The steps of the Gauss-Seidel algorithm are repeated until convergence (i.e., when $\bx^{(k+1)} = \bx^{(k)}$ and $\by^{(k+1)} = \by^{(k)}$) or a maximum number of iterations is reached.

\subsection*{Experimental Setup}

\emph{\textbf{Data Sources}}. 
Our case study is based on ride requests from the lower Manhattan area, in New York City, NY. We use real data recorded by the Taxi and Limousine Commission (TLC)\cite{TLC} on Tuesday, March 1, 2022, between 6:00 and 24:00. The total number of ride requests collected during this time window is 19935, from which we keep a random sample of 2480 to have a number of ride requests that can be handled by a fleet of 100 EVs. The ride requests have been categorized according to their submission time, and grouped into one-minute time slots. Note that the ride requests sub-sampling is due to the fact that we require to keep the simulation of several experiments manageable on a 2.4 GHz Quad-Core Intel Core i5 laptop with 8 GB RAM memory. The proposed algorithm on average converges in 2 iterations and takes 2.6 seconds per iteration.
We obtain approximated power generation profiles for PV systems from the renewable historical data of New York Independent System Operator\cite{NY_ISO_2023}. As a point of reference, we extracted the PV shape from the renewable energy generated within New York State on a sunny day, and modify it accordingly to represent also a cloudy day. To test our model, we consider three scenarios, corresponding to different weather conditions: sunny day (i.e., maximum PV generation), and cloudy day, which presents two alternatives: cloudy morning and cloudy afternoon. According to NYC OpenData\cite{PV}, we estimate that about 5 MW of PV peak power will be generated in the part of Manhattan that we are considering, distributed among a total of 39 charging stations. Assuming all PV systems have equal capacity, each station is estimated to produce about 125 kW during peak hours. Since most likely only a fraction of the generated power will be designated to EV charging, and given the smaller fleet size considered in this work, we assume that each charging station can produce 25 kW as peak power. The charging stations are grouped within the 4 regions they belong to by adding their PV capacity. In our graph, this translates to 4 nodes hosting charging facilities, as depicted in Figure \ref{fig:cartoonB}. Since the number of charging stations at each of these facilities varies, the maximum PV generation will be different in each region, as one can see in Figure \ref{fig:Bilevel_power4sation}.  

\vspace{.2cm}

\noindent
\emph{\textbf{Problem and algorithm setup}}. Next, we describe the parameters and functions used in the case study. We consider a fleet of 100 EVs, each with a 50 kWh battery capacity. 
We assume that each EV consumes 0.1~kWh per minute traveled, and recharges 0.2~kWh per minute spent at any charging station equipped with level 2 EV chargers, i.e., it takes approximately 4 hours to fully charge an empty EV battery. 
Based on our data sub-sample, it is reasonable to start our simulations with a fleet of idling EVs at 6:00, since there are very few ride requests and no renewable generation at night. However, the results do not change significantly even if we randomly set an initial state (idling, charging, driving) for each vehicle as long as the EVs become available in the next hour and a half.
To make our model more realistic, we implement some constraints summarized in the following. EVs whose battery status is up to $2/3$ of the full battery capacity cannot attend a charge request. EVs are expected to charge fully once they have accepted a charge request and will therefore not be considered for other assignments in the meanwhile. When they are available, EVs can only accept charge requests coming from the region where they are idling or from neighboring areas located at most one edge distance. EVs must have enough battery to complete the planned trip to be considered as candidates for a request. Ride requests can be attended by EVs located in the near neighborhood, i.e., no further than two edges distance. The travel time to move from one neighboring area to another is 10 minutes. The EV assignment is performed every $1$ minute.

The functions for the optimization problems are defined as follows:

$\bullet$  For the EV: $D_{ij}(y_{ij}, w_{ij}) = (y_{ij}-w_{ij})^2$, $\cY_\cR = [r_{\min}, r_{\max}], \, \forall i \in \cV, \, j \in \cR. $

$\bullet$   For the power utility company:
    $ U_s(x_{ij}, y_{ij}, \theta_s) = (L(P_{\mathrm{ref,s}}) - \rho_s(x_{ij}, y_{ij}))^2, \, \cY_\cC = [l_{\min}, l_{\max}],  \, \forall s \in \cS$, where
    $L(P_{\mathrm{ref,s}})  = c_{\mathrm{RER}} \left( P_{\mathrm{ref},s} - v_{\mathrm{ch},s} p_{\mathrm{ch} } \right)$, and $\rho_s(x_{ij}, y_{ij}) = \sum_{i \in \cV} \sum_{i \in \cC_s} x_{ij} y_{ij}, \, \forall s \in \cS$. 
    The function $L$ accounts for the loss incurred due to excess of renewables, where $P_{\mathrm{ref},s}$ is the power generated at specific areas where the charging facility $s$ is located, $v_{\mathrm{ch}}$ is the number of EVs currently charging at facility $s$, and $p_\mathrm{ch}$ is the power delivered to each charging EV. The time-varying price of the generated PV power is given by $c_{\mathrm{RER}}$. 

Numerically, if the EV $i$ cannot accommodate the ride request $j \in \cR$ or the charge request $j \in \cC$, we set $c_{ij} = 10^6$ or $d_{ij} = 10^6$, respectively; this means that the assignment is infeasible\cite{AS-LM-CG:19}. We remark that an alternative way to impose that the EV $i$ cannot accommodate the ride request $j$ is to set $y_{ij} = -10^6$. In a practical implementation of the algorithm, this represents the case where the EV driver $i$ cannot take the ride $j$ because the SOC is not sufficient (in this case, the ride-service provider has no knowledge of the SOC level of the EV).

\vspace{.2cm}

\begin{table}[h!]
\captionsetup{font={small,it},labelfont={bf,sf}}
\centering
\begin{tabular}{|l|l|l|}
\hline
\textbf{Variable Symbol} & \textbf{Description}  & \textbf{Value/Range} \\
\hline
$\cV, \; i \in \cV$ & set of EVs with cardinality $m$, index of EV & $m = 100$\\
\hline
$\cR, \; j \in \cR$ & set of ride requests with cardinality $p$, index of ride request & $p$ time-varying\\
\hline
$\cC, \; j \in \cC$ & set of charge requests with cardinality $q$, index of charge request & $q$ time-varying\\
\hline
$\cE = \cR \cup \cC$ & set of requests (ride and charge) & $|\cE|$ time-varying\\
\hline
$\cN$ & set of nodes with cardinality $n$ & $n = 9$ \\
\hline
$\cL \subseteq \cN \times \cN$ & set of edges & $|\cL| = 18$ \\
\hline
$\cS, \; s \in \cS \subseteq \cN$ & set of renewable-powered charging facilities, index of charging facility & $|\cS| = 4 $ \\
\hline
\hline
$\bbR^{m \times p+q} \ni \bX = [x_{ij}]$ & matrix of optimization variables for the vehicle-request assignment & $ x_{ij} \in \{0, 1\} $ \\
\hline
$\bx = \mathrm{vec}(\bX) = [\bx_{\cR}; \bx_{\cC}]$ & vector of optimization variables for ride $\bx_\cR$ and charge $\bx_\cC$ requests & $ x_{ij} \in \{0, 1\} $ \\
\hline
$\bbR^{m \times p+q} \ni \bY = [y_{ij}]$ & matrix of optimization variables for the (dis)incentive problem & $ y_{ij} \in \cY$ \\
\hline
$\by = \mathrm{vec}(\bY) = [\by_{\cR}; \by_{\cC}]$ & vector of optimization variables for EVs $\by_\cR$ and utility $\by_\cC$ incentives & $ y_{ij} \in \cY $ \\
\hline
$\bbR^{m \times p} \ni \bC = [c_{ij}]$ & matrix of operational cost for ride requests w.r.t. each EV & $ c_{ij} \in \bbR_{\geq 0}$ \\
\hline
$\bbR^{m \times q} \ni \bD = [d_{ij}]$ & matrix of operational cost for charge requests  w.r.t. each EV & $ d_{ij} \in \bbR_{\geq 0}$ \\
\hline
$\bbR^{m \times p} \ni \bW = [w_{ij}]$ & matrix of fixed (dis)incentives w.r.t. ride requests & $ w_{ij} \in \bbR$ \\
\hline
\hline
$P_{\mathrm{ref},s}$ & surplus of renewable energy at charging facility $s$ & $ P_{\mathrm{ref},s} \in \bbR_{\geq 0} $ \\
\hline
$v_{\mathrm{ch}}$ & number of EVs currently charging at facility $s$ & $v_{\mathrm{ch}} \in \bbR_{\geq 0}$  \\
\hline
$p_\mathrm{ch}$ & power delivered to each EV & $ p_\mathrm{ch} = 12$ kW \\
\hline
$c_{\mathrm{RER}}$ &  price kWh of the generated renewable power & time-varying\\
\hline
$h_j$ & user's bid for ride request $j$ & $ h_j \in [0, \, h_{\max}] $ \\
\hline
$\alpha_{ij}$ & weight of fixed incentive w.r.t. EV $i$ attending ride request $j$  & $\alpha_{ij} \in [0, 1]$\\
\hline
$a_{ij}$ & cost from EV $i$ location to request $j$ destination & $a_{ij} \in \bbR_{\geq 0}$\\
\hline
$\beta$ & weight of fixed incentive to ride-share  & $\beta \in [0, 1]$ \\
\hline
$b_i$ & fixed incentive to ride-share in EV $i$ & $b_i \in [0, 4]$ \\
\hline
$b_{\min, s}$, $b_{\max, s}$ & min/max for the total incentive at charging facility $s$ & $\{ b_{\min, s}, \, b_{\max, s} \} \in \bbR$  \\
\hline
$r_{\min}$, $r_{\max}$ & min/max for the (dis)incentive w.r.t. ride requests & $\{ r_{\min}, \, r_{\max} \} \in \bbR$  \\
\hline
$l_{\min}$, $l_{\max}$ & min/max for the incentive w.r.t. charge requests & $\{ l_{\min}, \, l_{\max} \} \in \bbR$  \\
\hline
\end{tabular}
\caption{\label{tab:Nomenclature} Definition of sets, variables, and parameters for our vehicle assignment method based on renewable and ride incentives.}
\end{table}

We remark that our numerical results consider the case where EVs charge until full once they accept a charging request. However, our mechanism can allow EVs to interrupt charging in order to accept a ride assignment; this is possible by  including the EVs that are charging in our Algorithm \ref{algo:Gauss-Seidel}  and, in particular, in the steps [S2]--[S3]. However, in a realistic implementation, the ability to interrupt charging will be regulated by contractual agreements associated with the charge request (for example, the power utility company may require a minimum charging time once the EV starts the charging session).

In terms of implementation of the Algorithm~\ref{algo:Gauss-Seidel}, it is important to remark that, although the assignment is performed in step [S3] by the ride-service provider, Algorithm~\ref{algo:Gauss-Seidel} is specifically designed for the EV drivers and the power utility company to influence the assignment process. As explained in the section Theoretical Foundations, Algorithm~\ref{algo:Gauss-Seidel} reaches a Nash equilibrium, where EV drivers have no incentive to deviate from the final assignment. For this reason, the numerical simulations considers a case where the EV drivers are fully compliant and serve the assigned ride or charge requests. Having said that, our algorithm is still applicable to a case where drivers are not fully compliant, and may decide to decline ride or charge requests that are assigned to them. These ride and charge requests  are re-dispatched in the next step or just missed.

\subsection*{Theoretical Foundations} 

In this section, we present two key concepts to analyze our vehicle assignment method based on renewable and ride incentives. 

\subsubsection*{Symmetric Linear Assignment Problem}

The problem in \eqref{eq:RHC_pro} is an asymmetric linear assignment problem\cite{Papadimitriou1998, Bijsterbosch2010}. By taking advantage of the totally unimodular constraint matrix property of the linear assignment problems in \eqref{eq:RHC_pro}, it is well known that their continuous relaxation (i.e., when one substitutes $x_{ij} \in \{0, 1\}$ with $x_{ij} \in [0, 1]$) is exact \cite{AS-LM-CG:19}. Therefore, we can reformulate the problem \eqref{eq:RHC_pro} by following a similar procedure as \cite{AS-LM-CG:19}. First, we add virtual requests $\cM_{E}$ or virtual EVs $\cM_{V}$ such that $\cE' = \cE \cup \cM_{E}$ and $\cV' = \cV \cup \cM_{V}$ have the same cardinality, and define $h~=~\max\{m, p+q\}$. Second, the assignment cost $c_{ij}$ (or $d_{ij}$) of the virtual EVs in $\cM_{V}$ for all requests in $\cM_{E}$ is set to $\infty$. The incentives for the virtual EVs are set to $0$. Thus, the problem \eqref{eq:RHC_pro} is equivalent to the following symmetric linear assignment problem: 
\begin{subequations} \label{eq:assig_2}
\begin{align} 
\min_{x_{ij}\in [0,1]} ~~ & \sum_{i=1}^h \sum_{j=1}^h (c_{ij} - y_{ij}) x_{ij} + \sum_{i=1}^h \sum_{j=1}^h (d_{ij} - y_{ij}) x_{ij}, \label{eq:assig_cost2} \\
\text{s.t.} ~~~~~& \sum_{i=1}^h x_{ij} = 1 \quad \forall j \in \cE', \label{eq:assig_c12} \\
& \sum_{j=1}^h x_{ij} = 1 \quad \forall i \in \cV'. \label{eq:assig_c22}
\end{align}
\end{subequations} 
We then have the following result. 

\begin{lemma}
Problem~\eqref{eq:assig_2} is an exact continuous relaxation of the binary Problem~\eqref{eq:RHC_pro}. 
\end{lemma}
The proof of the result follows from the totally unimodularity of the constraints, or equivalently from the fact that the solutions are vertices of the Birkhoff's polytope. This is a standard result in linear programming. The interested reader is referred to~\cite{Burkard1999,AS-LM-CG:19}.

\subsubsection*{Existence of the Nash Equilibrium}

The proposed method is aligned with game-theoretic frameworks\cite{facchinei_2010_NE-VI, Fukushima_1992, kubota2010gap, CO-GT-VI_2010}; in particular, it can be modeled as a non-cooperative game, where three groups of agents (ride-service provider, power utility company, and EVs) interact to optimize their costs. To analyze our method, we first consider the case where no constraints are present for the power utility company. To streamline the exposition, we define as $g_1(\bx_\cR; \by_\cR)$ and $g_2(\bx_\cC; \by_\cC)$
the cost functions in~\eqref{eq:PUC_pro} and~\eqref{eq:EVi_pro}, respectively, where the notation is provided in Table~\ref{tab:Nomenclature}.

\begin{definition}{(from\cite{facchinei_2010_NE-VI})} \label{def:GNE_def}
     A Nash Equilibrium (NE) is a tuple $(\bx^*, \by^*)$ such that 
    \begin{align*} 
    \bx^* &\in \arg \min_{\bx} ~ f(\bx, \by^*),  &  \by^* &\in \arg \min_{\by} ~ g(\bx^*, \by) := g_1(\bx_\cR^*, \by_\cR) + g_2(\bx_\cC^*, \by_\cC),  \\ 
    &\text{s.t.} ~~~ \bx \in \cX.                & &\text{s.t.} ~~~ \by \in \cY = \cY_\cR \times \cY_\cC. 
    \end{align*}
\end{definition}
Definition \ref{def:GNE_def} implies that when each agent chooses its strategy, $\bx^*$, $\by^*_\cR$ and $\by^*_\cC$, no one has any incentive to change its strategy unilaterally. Now, we reformulate the Nash Equilibrium problem (NEP) as a variational inequality (VI)\cite{facchinei_2010_NE-VI}. First, we define the vector-valued function $\map{F}{ \bbR^{2m(p+q)}}{ \bbR^{2m(p+q)}}$ and the point-to-set mapping $S: \bbR^{2m(p+q)} \rightrightarrows \bbR^{2m(p+q)}$ by
    \begin{align} \nonumber
        F(\bx, \by) &= \begin{bmatrix}
        \nabla_{x} f(\bx, \by) \\
        \nabla_{y} g(\bx, \by)
        \end{bmatrix} 
        \quad \text{ and } \quad S := \cX \times \cY
    \end{align}
 Therefore, we can define the VI problem as follows.
 \begin{definition} (from\cite{facchinei_2010_NE-VI})
     Given a closed and convex set $S \subseteq \bbR^{2m(p+q)}$ and a mapping $\map{F}{S}{\bbR^{2m(p+q)}}$, the VI problem, denote by $VI(S,F)$, consists in finding a vector $\bz^* := [\bx^*, \by^*]^\top$, such that:
        $(\bh - \bz^*)^\top F(\bz^*) \geq 0,  \forall~\bh \in S$.
 \end{definition}

 In\cite{facchinei_2010_NE-VI}, the relation between NEP and VI is established. In particular, suppose that: 1) The set $S$ is convex and closed. 2a) The cost functions $f(\bx, \by)$ is continuously differentiable in $(\bx, \by)$ and convex in $\bx$ for every fixed $\by$, and 2b) The cost functions $g(\bx, \by)$ is continuously differentiable in $(\bx,\by)$ and convex in $\by$ for every fixed $\bx$. Then, the NEP is equivalent to the $VI(S, F)$.
 The conditions that guarantee the existence of a NE follow directly from the existence of a solution of the VI. To show the existence, in addition to conditions 1) and 2), we have that, in our case, the set $S$ is compact; consequently, the NEP has a nonempty and compact solution set\cite{facchinei_2010_NE-VI}. 

When one considers the constraints in problem~\eqref{eq:PUC_pro}, we need to resort to the concept of Generalized Nash Equilibrium. Based on the  problems~\eqref{eq:RHC_pro}--\eqref{eq:EVi_pro}, define
$\hat{S}_x = \cX, \; \hat{S}_y(\bx) = \{ \by \in \cY \, | \, {b}_{\min} \leq \rho_s(\bx, \by) \leq {b}_{\max}, \, \forall s \in \cS \}$ and $\hat{S} = \hat{S}_x \times \hat{S}_y(\bx)$.
Then, the NEP becomes a Generalized NEP (GNEP) where the set $\hat{S} \subseteq \bbR^{2m(p+q)}$ depends on the other agents’ strategies. Assume $\bH$ is a given symmetric positive definite matrix, and let $\bz$ be temporarily fixed. Consider the following problem\cite{kubota2010gap, Fukushima_1992}:
\begin{equation} \label{eq:aux1}
    \min_{\bh \in \hat{S}} \left\{ \inner{F(\bz), \, \bh-\bz} + \frac{1}{2} \inner{\bh-\bz, \, \bH(\bh -\bz)}  \right\} \quad \equiv \quad \min_{\bh \in \hat{S}} \| \bh - (\bz - \bH^{-1} F(\bz)) \|_\bH^2,
\end{equation}
and the solution of the problem \eqref{eq:aux1} is given by $h(\bz) = \mathrm{proj}_{\hat{S}, \bH} \{\bz - \bH^{-1} F(\bz)\}$,
where the mapping $\map{h}{\bbR^{2m(p+q)}}{\bbR^{2m(p+q)}}$ yields a fixed point characterization of the solution of the quasi-variational inequality (QVI) problem, defined as $(\bh - \bz^*)^\top F(\bz^*) \geq 0, \, \forall \bh \in \hat{S}$. We can define the merit function 
of the GNEP as the function $u(\bz)$ defined as \cite{Fukushima_1992} $u(\bz) = - \inner{F(\bz), \, h(\bz)-\bz} + \frac{1}{2} \inner{h(\bz)-\bz, \, \bH(h(\bz) -\bz)}$, and consider the problem $\min_{\bz} \; u(\bz), s.t.~\bz \in \hat{S}$.  

From\cite{Fukushima_1992}, it follows that for each $\bz \in \hat{S}$, we have $u(\bz) \geq 0$; moreover, $\bz$ solves the QVI problem if and only if $u(\bz) = 0$ and $\bz \in \hat{S}$. This result gives a tool to corroborate numerically the performance of Algorithm \ref{algo:Gauss-Seidel}. In particular, if the assignment and the incentives at convergence are such that the merit function is $0$, then the proposed Gauss-Seidel method identified a GNE.

\section*{Data and Code Availability}

Data for the ride requests was obtained from the Taxi and Limousine Commission (TLC), publicly available on the website: \href{https://www.nyc.gov/site/tlc/about/tlc-trip-record-data.page}{https://www.nyc.gov/site/tlc/about/tlc-trip-record-data.page}. All analyses were performed using Python software that was custom developed as part of our methods. It is available in the repository: \href{https://github.com/Elispe/renewableEV}{https://github.com/Elispe/renewableEV}. Here we deploy the open source CVXPY \cite{diamond2016cvxpy} package for convex optimization problems and PuLP \cite{pulp} for the linear assignment problem, with the COIN\_CMD solver.

\bibliography{References}


\section*{Acknowledgements}
The work of E. Perotti was supported by Schmidt Science Fellows, in partnership with the Rhodes Trust. The work of A. Ospina and E. Dall'Anese was supported in part by the  National Science Foundation (NSF) award 1941896 and by the NSF ERC ASPIRE.

\section*{Author contributions statement}

E.D. and G.B. conceived the research, all authors contributed to the methods,  A.O. and E.P. conducted the experiments, A.S. supervised the technical developments. All authors analyzed the results and reviewed the manuscript. 

\section*{Additional information}
 The authors declare no competing interests. 


\end{document}